\documentclass[12pt]{article}
\usepackage{amssymb}
\usepackage{amsmath}
\usepackage{graphicx}

\catcode`@=11
\renewcommand{\section}{\@startsection{section}{1}{0pt}{\medskipamount}
{\medskipamount}{\large\bf}}
\numberwithin{equation}{section}
\catcode`@=12

\def\beq{\begin{eqnarray}}    
\def\eeq{\end{eqnarray}}      

\def\Tr{\,\mbox{Tr}\,}                  
\def\sDet{\,\mbox{sDet}\,}              

\def\pa{\partial}                       

\def\im{\textrm{i}}

\def\diff{\textrm{d}}
\def\sfrac#1#2{{\textstyle\frac{#1}{#2}}}
\def\={\ =\ }

\def\be{\beta}

\def\de{\delta}

\def\vp{\varepsilon}

\def\ph{\varphi}

\def\Ga{\Gamma}

\def\La{\Lambda}
\def\Si{\Sigma}

\begin{document}

\begin{titlepage}
\setcounter{page}{0}

\vskip 2.0cm

\begin{center}

{\large\sc On the Functional Renormalization Group approach for
Yang-Mills fields}
\vspace{10mm}

{\large Peter M. Lavrov $^{a,b}$ \ \ and \ \
Ilya L. Shapiro $^{b,a}$}

\vspace{6mm}

\noindent ${}^a${\em
Tomsk State Pedagogical University,\\
Kievskaya St.\ 60, 634061 Tomsk, Russia}
\vspace{4mm}

$^b${\em  Departamento de F\'{\i}sica, ICE, 
Universidade Federal de Juiz de Fora,
\\
Juiz de Fora, CEP: 36036-330, MG,  Brazil}
\vskip 12mm

\begin{abstract}
\noindent
We explore the gauge dependence of the effective average action
within the functional renormalization group (FRG) approach. It is
shown that in the framework of standard definitions of FRG for
the Yang-Mills theory, the effective average action remains
gauge-dependent on-shell, independent on the use of truncation
scheme. Furthermore, we propose a new formulation of the FRG,
based on the use of composite operators. In this case one can
provide on-shell gauge-invariance for the effective average
action and universality of $S$-matrix.
\vskip 6mm

\noindent {\sl Keywords:} \ \
Renormalization group, \ Effective Average Action,
\ Yang-Mills theories, \, Gauge dependence,
\ BRST symmetry, \ Slavnov-Taylor identity, \
Composite fields.
\vskip 2mm

\noindent {\sl PACS:} \
11.10.Gh, \ 
11.10.Hi, \ 
11.15.-q, \ 
11.15.Bt, \ 
11.15.Tk  \ 
\vskip 2mm

\noindent
{\sl MSC-AMS:} \
81T15, \ 
81T13  \ 
81T16  \ 
81T17  \ 

\end{abstract}
\end{center}
\hfill

\noindent{\sl E-mails:} \ lavrov@tspu.edu.ru, \quad shapiro@fisica.ufjf.br

\end{titlepage}

\section{Introduction}

\noindent
The recent development of Quantum Field Theory is greatly related
to the non-perturbative aspects of quantum theories. The request
for such a non-perturbative treatment is related to the triviality
problem of scalar field theory, non-perturbative nature of
low-energy QCD and also an expectation to achieve a consistent
theory of Quantum Gravity. One of the most promising approaches
is related to different versions of Wilson renormalization
group approach \cite{Wilson}. An important advance towards the
use of the non-perturbative renormalization group has been done
in the paper \cite{Polch}. The qualitative idea of this work can
be formulated as follows: regardless we do not know how to sum up
the perturbative series, in some sense there is a good qualitative
understanding of the final output of such a summation for the
propagator of the quantum field. An exact propagator is supposed
to have a singe pole and also provide some smooth behavior in both
UV and IR regions. It is possible to write a cut-off dependent
propagator which satisfies these requirements. Then the cut-off
dependence of the vertices can be established from the general
scale-dependence of the theory which can be established by means
of the functional methods. The method proved to be very helpful,
in particular, for understanding the perturbative renormalization
of the theory.

A compact and elegant formulation of the non-perturbative
renormalization group has been proposed in \cite{{Wett-1,{Wett-2}}}
in terms of effective action. The method was called functional
renormalization group (we shall use abbreviation FRG) for the
effective average action, it is nowadays one of the most popular and
developed QFT methods, which can be seen from the review papers on
the FRG method \cite{FRG1,FRG2,FRG3,FRG4,IIS-2009,FRG5,FRG6}.

As far as some of important applications of the FRG approach is
related to QCD and Quantum Gravity, the special attention has been
paid to the study of effective average action in gauge theories
\cite{Wett-Reu-1,Wett-Reu-2} (see also
\cite{FRG7,Bob-Att-Mar-1,Ell-1,Att-TimM,FRG4,Lit-Paw-1,{ReWe-1997},
Fre-Lit-Paw,Iga-Ito-So} and a very clear and complete review
\cite{Giess}). Many aspects of gauge theories in the framework of
FRG has been discussed with success, but there is still one important
question which remains unsolved. The consistent quantum description
of gauge theories has to provide the on-shell independence on the
choice of the gauge fixing condition. In a consistent formulation,
such an independence should hold for the $S$-matrix elements and,
equivalently, for the on-shell effective action. There is a good
general understanding that this point represents a difficulty for
the FRW approach
\cite{Wett-Reu-2,Bob-Att-Mar-1,Ell-1,Att-TimM,Lit-Paw-1,Iga-Ito-So,Giess},
because the construction of FRG starts from the propagator, which is
not a gauge invariant object and, in particular, always depends on
the choice of gauge fixing condition. However, as far as we could
see, the complete analysis of whether this general difficulty leads
to problems at the level of $S$-matrix, was not done. The first
purpose of the present work is to fill this gap, so we present
a formal consideration of gauge dependence for the case of pure
Yang-Mills theory. The method of analysis
employed here is based on a standard (albeit not really simple) use
of BRST symmetry \cite{brst} which plays a fundamental role in
Quantum Gauge Field Theory \cite{books}, and local form of
Slavnov-Taylor identities \cite{S,T}. As we shall see in what
follows, the regulator functions which emerge in the modified
propagators do violate BRST symmetry and this leads, eventually, to
the on-shell gauge dependence of the effective average action and,
consequently, to the ambiguous $S$-matrix.

The existing attempts to solve the problem of gauge invariant
formulation of FRG can be classified into two different types.
The first one is based on reformulating the Yang-Mills theory
with the help of a gauge-invariant cut-off-dependent regulator
function introduced as a covariant form factor into the action
of Yang-Mills fields, so that the regulated action is gauge
invariant \cite{TimM-JHEP}, \cite{Rost} (see also earlier work
\cite{MorKub}). Then the renormalization group equation is
formulated. As far as there are no gauge fixing, one is free
from the ambiguity related to the choice of the gauge condition.
It is supposed that the infinite integral over the gauge group is
absorbed into vacuum functional renormalization. It is not clear
for us to which extent this approach for implementing covariant
cut-off has relation to the effective average action of
\cite{Wett-Reu-1,Wett-Reu-2}. An obvious deviation from the
``canonical'' method is that inserting the covariant form factor
into the action of Yang-Mills fields means that the vertices also
become cut-off dependent. According to \cite{Giess}, from the
viewpoint of applications this, very interesting, approach
requires dealing with complicated non-local structures. Also,
the divergences which remain after integrating over the gauge
group in the non-Abelian theory can depend on the Yang-Mills
fields and, therefore, their removal without usual
renormalisation procedure may be a difficult task.

The second approach \cite{Ven,Paw-2} is based on the use of
Vilkovisky unique effective action \cite{Vilk-84} (see also
\cite{DeWitt-87} and \cite{FrTs-EdEA-84} for further developments).
The unique effective action provides gauge independence not only
for the $S$-matrix, but even for the off-shell effective action.
The price one has to pay is that this construction has its own
ambiguities connected.  It would be certainly interesting to
have an alternative formulation of the effective average action,
which would possess, in part of gauge dependence, the same
properties as the conventional effective action in QFT.
Namely, it may be gauge dependent off-shell, but should be
gauge independent on-shell, such that the $S$-matrix would be
unitary and well-defined.

As far as the source of the problem with gauge non-invariance is
the introduction of the cut-off (or, better say, scale-dependent)
propagator, it is clear that this is the aspect of the theory
which should be reconsidered first. The known theorems about
gauge-invariant renormalizability \cite{VLT-82,GomWein} tell us that
the exact effective action should be BRST-invariant. As far as the
regulator functions which modify the propagator are supposed to
mimic the all-loop quantum corrections, they must be taken in the
BRST-invariant form. Therefore, the problem is just to find the
way to implement this invariance when one takes into account the
regulator functions. The proposal which we present here is to use an
old idea of \cite{L1} about introduction of composite operators in
gauge theories. Following this line, we will introduce the regulator
functions as composite operators and show that, in this case, the
BRST symmetry is maintained at quantum level and, as a consequence,
the $S$-matrix in the theory is gauge invariant and well defined.

The paper is organized as follows. The main features of the
Faddeev-Popov method \cite{FP} for Yang-Mills fields are described
in Section~2. For the pedagogical purposes we also include the
demonstration of the gauge invariance of the vacuum functional
and the on-shell invariance of effective action. Let us note that
the rest of the paper relies on this important section in many
respects, including definitions and notations. In Section~3, the
FRG approach for Yang-Mills fields \cite{Wett-2,Wett-Reu-2} is
briefly reviewed. In Section~4, the gauge dependence of vacuum
functional in the FRG approach is investigated. In Section~5,
the gauge dependence of effective average action of the FRG
approach is explored. In Section 6 we present the new approach
to the quantization of Yang-Mills fields with the regulator
functions introduced by means of composite operators. Finally,
Section 7 consists of concluding remarks and final discussions.

We use the standard condensed notation of DeWitt~\cite{DeWitt}.
Derivatives with respect to sources and antifields are taken
from the left, while those with respect to fields are taken
from the right. Left derivatives with respect to fields are
labeled by a subscript~$l$. The Grassmann parity of a quantity
$F$ is denoted as $\vp(F)$.

\section{ Yang-Mills theories within the Faddeev-Popov quantization}

\noindent
In this section we shall present some basic facts about Yang-Mills
fields within the Faddeev-Popov quantization method \cite{FP}. Up
to some extent, these considerations are general, but we restrict
our attention to the Yang-Mills case only, just to stay within the
scope of the present work. Our main purpose is to discuss the
gauge independence of vacuum functional and, consequently, the
gauge independence of the generating functional of the vertex
functions (effective action) on-shell. Despite this material has
not been presented earlier in exactly this form, the section has
introductory purpose, and is intended to serve as a reference for
the consequent consideration of the same issues in the framework
of functional renormalization group approach to Yang-Mills
theory which will be dealt with in the next sections.

The Yang-Mills fields $A^a_{\mu}(x)$ belong to the adjoint
representation of the $SU(n)$ group, such that $a=1,\ldots,n^2{-}1$.
The initial classical action $S_0$ has the standard form,
\beq
S_0(A) \= -\frac{1}{4}\int\!\diff^D x\ F_{\mu\nu}^{a}F^{\mu\nu\,a}
\,,
\qquad\textrm{with}\qquad
F^a_{\mu\nu}\=\partial_{\mu}A^a_{\nu}-\partial_{\nu}A^a_{\mu}
+ f^{abc}A^b_{\mu}A^c_{\nu}\,,
\label{clYM}
\eeq
where $\mu,\nu=0,1,\ldots,D{-}1$ and $f^{abc}$ denote the (totally
antisymmetric) structure constants of the Lie algebra~$SU(n)$.
We assume that Minkowski space has the signature $(-,+,\,...\,,+)$.
The action (\ref{clYM}) is invariant under gauge transformations
\beq
\label{gs}
\delta A^a_{\mu}\=D^{ab}_{\mu}\xi^b
\qquad
\mbox{with}
\qquad
D^{ab}_{\mu}\=\delta^{ab}\partial_{\mu}+f^{acb}A^c_{\mu}
\eeq
being generators of these transformations. After the Faddeev-Popov
quantization, the field configuration space of Yang-Mills theory
\beq
\Phi^A\,=\,\{A^a_\mu, C^a, {\bar C}^a, B^a\}\,,
\quad \mbox{with} \qquad
\varepsilon(C^a)=\varepsilon(\bar C)^a=1
\,,\quad \varepsilon(A^a_\mu)=\varepsilon(B^a)=0
\eeq
includes the (scalar) Faddeev-Popov ghost and antighost fields
\ $C^a$ \ and \ ${\bar C}^a$, respectively, as well as the
Nakanishi-Lautrup auxiliary fields $B^a$. Choosing gauge fixing
condition
\beq
\chi^a(A,B)=0\,,
\label{gf}
\eeq
the Faddeev-Popov action, $S_{FP}$, is constructed in the form
\beq
\label{FPact}
S_{FP}(\Phi)=S_0(A)\ +\ {\bar C}^aM^{ab}(A,B)C^b \,+\,
\chi^a(A,B)B^a \,,
\eeq
with
\beq
M^{ab}(A,B)
&=&
\frac{\de\chi^a(A,B)}{\de A^c_{\mu}}\,D^{cb}_{\mu}\,.
\eeq
The most popular gauge functions \ $\chi^a$ \ in the Yang-Mills
theory are the Landau gauge,
\beq
\chi^a=\partial^{\mu}A_{\mu}^a\,,
\eeq
and the \ $R_{\xi}$ \ gauge, defined by
\beq
\chi^a=\partial^{\mu}A_{\mu}^a+\frac{\xi}{2}\,B^a\,,
\eeq
where $\xi$ is an arbitrary gauge parameter. For these two cases
the Faddeev-Popov matrices \ $M^{ab}$ \ have the same form
\beq
M^{ab}=\partial^{\mu}D^{ab}_{\mu}\,.
\eeq

The action (\ref{FPact})  is invariant under the BRST transformation
\cite{brst}
\beq
\label{BRSTtr}
\delta_B A_{\mu}^{a} = D^{ab}_{\mu}C^b\theta
\,,\qquad
\delta_B \bar{C}^a = B^a\theta
\,,\qquad
\delta_B B^a = 0
\,,\qquad
\delta_B C^a = \sfrac12 f^{abc}C^bC^c\theta\,,
\eeq
where $\theta$ is a constant Grassmann parameter. This
transformation possesses a very important property of
nilpotency. Let the BRST transformation be presented
in the form
\beq
\delta_B\Phi^A \,=\, {\hat s}\Phi^A\theta
\,,\qquad
\varepsilon(\Phi^A)=\varepsilon_A\,,
\label{s-nil}
\eeq
then one can verify the nilpotency of BRST transformation
\beq
\label{BRSTnil}
{\hat s}^2 A_{\mu}^{a} ={\hat s}
D^{ab}_{\mu}C^b=0
,\quad
{\hat s}^2 \bar{C}^a = {\hat s}B^a=0
,\quad {\hat s}^2 B^a = 0
,\quad
{\hat s}^2C^a = {\hat s}\sfrac12
f^{abc}C^bC^c =0.
\eeq
The generating functional of the Green's functions, which is
sufficient to calculate all processes with Yang-Mills fields
is given by the Faddeev-Popov formula \cite{FP}
\beq
\label{Zj}
Z(j)=\int {\cal D}{\Phi}\;{\rm exp} \Big\{
\frac{i}{\hbar}\big[S_{FP}({\Phi})+jA\big]\Big\}\,,
\eeq
where \ $j=\{j^a_{\mu}(x)\}$ \ are sources of the fields
\ $A=\{A^{a\mu}(x)\}$.

Thanks to the gauge invariance of the Yang-Mills action (\ref{clYM})
and to the BRST invariance of the extended action (\ref{FPact}), the
Green's functions of the theory obey the relations known as the
Slavnov- Taylor identities \cite{S,T}. These identities can be
derived  from (\ref{Zj}) by means of the change of integration
variables $A^a_{\mu}$, in the form of infinitesimal gauge
transformations (\ref{gs}). The Jacobian of these transformations is
equal to unity. Then the basic Slavnov-Taylor identities for
Yang-Mills fields can be written in the form
\beq
\label{STorig}
j^a_{\mu} \langle D^{\mu ab} \rangle_j \,+\, \langle
B^a\partial^{\mu}D^{ab}_{\mu}\rangle_j
\,+\, f^{acb} \langle {\bar C}^a\pa^\mu D^{cd}_\mu C^d
\rangle_j \,\equiv \,0\,,
\eeq
where the
symbol \ $\langle G \rangle_j$ \ means vacuum expectation value of
the quantity \ $G$ \ in the presence of external sources \
$j^a_{\mu}$,
\beq
\nonumber
\langle G \rangle_j \,=\, \int {\cal
D}{\Phi}\,\, G \, \exp \Big\{
\frac{i}{\hbar}\big[S_{FP}({\Phi})+jA\big]\Big\}\,.
\eeq
The generating functional \ $Z(j)$ \ contains information about all
Green's functions of the theory, which can be obtained by taking
variational derivatives with respect to the sources. Similarly, the
Slavnov-Taylor identities represent an infinite set of relations
obtained from (\ref{STorig}) by taking derivatives with respect
to external sources \ $j^a_{\mu}$.

The form of the Slavnov-Taylor identities can be greatly simplified
by introducing extra sources to the ghost, antighost and auxiliary
fields. In this case one has to deal with the extended generating
functional of the theory
\beq
\label{Z}
Z(j,{\bar\eta},\eta, \sigma)\,=\,
\int {\cal D}{\Phi} \;{\rm
exp} \Big\{\frac{i}{\hbar}\big[S_{FP}({\Phi})+jA+{\bar\eta}
C+\eta{\bar C}+\sigma B\big]\Big\}\,.
\eeq
It is clear that the relation to the conventional generating
functional (\ref{Zj}) performs as follows:
\beq
Z(j)=Z(j,{\bar\eta},\eta, \sigma)\Big|_{\eta={\bar\eta}=\sigma=0}\,.
\eeq

The generating functional of connected Green's functions,
$W(j,{\bar\eta},\eta, \sigma)$, is defined by the relation
\beq
Z(j,{\bar\eta},\eta, \sigma)={\rm exp} \Big\{
\frac{i}{\hbar}\,W(j,{\bar\eta},\eta, \sigma) \Big\}\,.
\label{W}
\eeq
Finally, the generating functional of vertex Green's functions
(effective action) is defined through the Legendre transformation
of \ $W$,
\beq
\label{GI}
\Gamma(A, C, {\bar C}, B)=W(j,{\bar\eta},\eta,\sigma)
-jA-{\bar\eta} C-\eta{\bar C}-\sigma B\,,
\eeq
where the source fields \ $j,\,{\bar\eta},\,\eta,\, \sigma$ \
are solutions of the equations
\beq
\label{GII}
A^{a\mu}(x)=\frac{\delta W}{\delta j^a_{\mu}(x)},\quad
C^a(x)= \frac{\delta W}{\delta {\bar\eta}^a(x)},\quad {\bar
C}^a(x)=\frac{\delta W}{\delta \eta^a(x)},\quad B^a(x)=\frac{\delta
W}{\delta \sigma^a(x)}.
\eeq
By means of (\ref{GI}) and (\ref{GII}) one can easily arrive at
the relations
\beq
\nonumber
&&\frac{\delta \Gamma}{\delta A^{a\mu}(x)}=-j^a_{\mu}(x)
\,,\qquad
\frac{\delta \Gamma}{\delta
B^a(x)}=-\sigma^{a}(x),
\\
\label{GII-bis}
&&\frac{\delta \Gamma}{\delta
{C}^a(x)}=-{\bar\eta}^a(x)
\,,\qquad
\frac{\delta \Gamma}{\delta {\bar C}^a(x)}=-\eta^a(x)\,.
\eeq

The functional $\Gamma$ satisfies the following functional
integro-differential equation
\beq
\label{GEq}
\exp {\Big\{}\frac{i}{\hbar}\,\Gamma(\Phi){\Big\}}
\,=\,\int {\cal D}\varphi \,\,
\exp \Big\{ \frac{i}{\hbar}\,\Big[
S_{FP}({\Phi +\varphi})
- \frac{\delta\Gamma(\Phi)}{\delta\Phi}\varphi \Big]{\Big\}}\,.
\eeq
The last equation is a good starting point to perform the loop
expansion, corresponding representation in form of the series in
\ $\hbar$,
\beq
\label{GEloop}
\Gamma(\Phi) \,=\, \sum\limits_{k=0}^\infty
\hbar^k\,\Gamma^{(k)}(\Phi)\,.
\eeq
The solution can be immediately found in the tree approximation,
\beq
\nonumber
\Gamma^{(0)}(\Phi)=S_{FP}(\Phi)\,.
\eeq

The Slavnov-Taylor identities which are consequences of gauge
symmetry of initial action can be rewritten with the help of
BRST symmetry of the Faddeev-Popov action. For this end we make
use of the change of variables in the functional integral (\ref{Z})
of the form (\ref{BRSTtr}). Because of the antisymmetry property of
structure coefficients \ $f^{abc}$ \ and nilpotency of \ $\theta$,
the Jacobian of this transformation is equal to 1. Using the
invariance of the functional integral under change of integration
variables, the following identity holds
\beq
\label{WIZI}
\int {\cal D}\Phi
\,\big(j\delta_B A+{\bar\eta}\delta_B C + \eta\delta_B{\bar C}\big)
\, \exp \Big\{ \frac{i}{\hbar}\,\big[S_{FP}(\Phi)
+ jA + {\bar\eta} C + \eta{\bar C} + \sigma B\big] \Big\}
\,\equiv\, 0\,.
\eeq
Here the nilpotency of BRST transformation and the
consequent exact relation
\beq
\label{exp}
\exp \Big\{ \frac{i}{\hbar}\,
\big(j\delta_B A + {\bar\eta}\delta_B C
+ \eta\delta_B {\bar C}\big)\Big\}
\,=\,
1 + \frac{i}{\hbar}\,\big(j\delta_B A+{\bar\eta}\delta_B
C+\eta\delta_B{\bar C}\big)
\eeq
has been used.

Using the invariance (\ref{WIZI}), one can easily arrive at the
Slavnov-Taylor identity for \ $Z$,
\beq
\label{WIZII}
j^a_{\mu}\,\pa^\mu\,\frac{\delta Z}{\delta{\bar\eta}^a}
\,+\,
\eta^a\,\frac{\delta Z}{\delta\sigma^a}
\,+\,
\frac{\hbar}{i}\,f^{acb}\,\Big(J^a_{\mu}
\,\frac{\de^2 Z}{\de j^c_{\mu}
\de{\bar\eta}^b}
\,+\,
\frac{1}{2}\,{\bar\eta}^a
\,\frac{\de^2 Z}{\de {\bar\eta}^c \de{\bar\eta}^b}
\Big)\,\equiv\, 0\,.
\eeq
Due to the presence of the second-order variational derivatives
of \ $Z$, the last identity has a non-local form. Fortunately,
there exists a possibility to present the Slavnov-Taylor identity
in the local form using the Zinn-Justin trick \cite{Z-J}. For the
sake of symmetry and compactness of notations, we introduce the
set of sources
\beq
J_A=(j^a_{\mu}, \,{\bar \eta}^a,\,\eta^a,\sigma^a)
\,,\qquad
\vp(J_A)\,=\,\vp(\Phi^A)\,=\,\vp_A\,,
\nonumber
\eeq
the set of external sources
\beq
K_A=(K^a_{\mu},{\bar L}^a, L^a, N^a)\,\qquad
\vp(K_A)=\vp_A+1
\nonumber
\eeq
to the BRST transformation,
\ ${\hat s}\Phi^A$, and the extended generating functional of
Green's functions
\beq
\label{Zext}
{\cal Z}( J,K)=\int {\cal D}{\Phi}\;
\exp \Big\{ \frac{i}{\hbar}\,\big[S_{FP}(\Phi) + J\Phi
+ K{\hat s} \Phi\big]\Big\}\,,
\eeq
where we used the notation for BRST transformations, which
was previously introduced in (\ref{s-nil}). It is clear that
\beq
\label{ZJK}
{\cal Z}(J,K)\Big|_{K=0}
\,=\, Z(j,{\bar\eta},\eta, \sigma).
\eeq
Making use of the change of variables (\ref{BRSTtr}) in Eq.
(\ref{ZJK}) and taking into account the nilpotency of BRST
transformation (\ref{BRSTnil}), we obtain
\beq
\label{WIZJK}
\int {\cal D}\Phi\,\,J_A \,{\hat s}\Phi^A\,\,
\exp \Big\{ \frac{i}{\hbar}\,\big[S_{FP}({\Phi})
+ J\Phi + K {\hat s}\Phi\big]\Big\}\,\equiv\,0
\eeq
or, equivalently,
\beq
\label{WIZJKI}
J_A\,\frac{\de {\cal Z}(J,K)}{\de K_A}\,\equiv\, 0\,.
\eeq
The last relation (\ref{WIZJKI}) represents the Slavnov- Taylor
identity for Yang-Mills theory in the local form.

The Faddeev-Popov quantization of Yang-Mills theories provides
a very important property of physical $S$-matrix being gauge
independent. Let us discuss this aspect of the theory. As a first
step, consider the vacuum functional
\ ${\cal Z}(0)\equiv {\cal Z}_{\chi}$ \
constructed  for a given choice of gauge $\chi^a=0$,
\beq
\label{WIZv}
{\cal Z}_{\chi}\,=\,
\int {\cal D} \Phi\,\,\exp
\Big\{\frac{i}{\hbar}\,S_{FP}(\Phi)\Big\}\,.
\eeq
Consider an infinitesimal change of the gauge fixing function
\ $\chi^a\rightarrow \chi^a+\delta\chi^a$, corresponding to the
new gauge fixing condition \ $\chi^a+\delta\chi^a=0$. We have
\beq
\label{WIZvd}
{\cal Z}_{\chi+\delta\chi} \,=\,
\int {\cal D}\Phi\,
\exp \Big\{\frac{i}{\hbar}\,\Big[S_{FP}({\Phi})
+ {\bar C}^a\,\frac{\de\,\de\chi^a}{\de A^c_\mu}\,D^{cb}_{\mu}C^b
+ \de\chi^a B^a\Big]\Big\}\,.
\eeq

Let us perform the change of variables according to Eqs.~(\ref{BRSTtr})
in the functional integral (\ref{WIZvd}), but with a functional
$\Lambda=\Lambda(\Phi)$ instead of the constant Grassmann
odd variable $\theta$. Here \ $\Lambda(\Phi)$ \ is supposed to
be a Grassmann-odd quantity. Of course, the Faddeev-Popov action,
\ $S_{FP}$, \ is invariant under such change of variables. The
contributions come only from the integration measure, resulting in
the corresponding Jacobian. Restricting our attention to the terms
of the first order in the Grassmann-odd quantity \ $\Lambda(\phi)$
\ and in the small quantity \ $\delta\chi_{\alpha}(A)$, one can
rewrite the Jacobian according to the usual relation \
$\sDet (I + M) = \exp ( {\rm sTr} M)$,
where \ ${M^A}_B \equiv {\delta (\delta\Phi^A)}/{\delta\Phi^B}$.
In this way we arrive at
\beq
\label{FPVac1}
{\cal Z}_{\chi + \delta\chi}
&=&
\int {\cal D}\Phi\,
\exp\bigg\{\frac{i}{\hbar}\,\Big[
S_{FP}(\Phi)
+ {\bar C}^a\frac{\de\de\chi^a}{\delta A^c_\mu} D^{cb}_{\mu}C^b
+ \delta\chi^aB^a
\\
\nonumber
&+&
i\hbar\, \frac{\de\La}{\delta A^a_\mu}\,D^{ab}_\mu\,C^b
\,-\, \frac{i\hbar}{2}\, f^{abc}\,C^b C^c\,\frac{\de\La}{\de C^a}
\,+\,i\hbar\,\frac{\de\La}{\de {\bar C}^a}\,B^a \Big]\bigg\}\,.
\eeq
By choosing the functional \ $\La(\phi)$ \ according to
\beq
\nonumber
\Lambda=\frac{i}{\hbar}\bar{C}^a\delta\chi^a\,,
\eeq
it is easy to see from (\ref{FPVac1}), that the vacuum functional does
not depend on the choice of gauge, namely
\beq
{\cal Z}_{\chi + \delta\chi}\,=\,{\cal Z}_{\chi}\;.
\label{Inva-Vac}
\eeq
Starting from this relation, one can prove the gauge
independence of the $S$ - matrix \cite{KT,Tyutin}.

The next part of our consideration concerns gauge independence
on-shell for the effective action. We start by introducing the
generalized generating functional of connected Green's functions
\begin{eqnarray}
\label{Wd}
{\cal Z}(J,K)\,=\, \exp \Big\{\frac{i}{\hbar}\,
{\cal W}(J,K) \Big\}\,.
\end{eqnarray}
Now one can rewrite the Slavnov-Taylor identity (\ref{WIZJKI}) in
terms of ${\cal W}$ as
\beq
\label{WIWJKI}
J_A\frac{\delta {\cal W}(J,K)}{\delta K_A} \,\equiv\, 0\,.
\eeq
The effective action, $\Gamma$, is introduced through the Legendre
transformation of ${\cal W}$,
\beq
\label{Gd}
\Gamma(\Phi, K)={\cal W}(J,K)-J_A\Phi^A
\,,\quad \mbox{where}\quad
\Phi^A \,=\,\frac{\delta {\cal W}}{\delta J_A}\,.
\eeq
From (\ref{Gd}) it follows that
\beq
\label{Gdpr}
\frac{\de \Ga}{\de \Phi^A} = - J_A
\quad \mbox{and}\quad
\frac{\de \Ga}{\de K_A} = \frac{\de {\cal W}}{\de K_A}\,.
\eeq

By performing a shift of the integration variable, one can show
that the functional $\Gamma$ satisfies the equation
\beq
\label{Gexteq}
\exp \Big\{ \frac{i}{\hbar}\,\Ga(\Phi,K)\Big\}
\,=\,\int {\cal D}{\varphi}\,\exp \Big\{\frac{i}{\hbar}\,\Big[
S_{FP}({\Phi+\varphi}) - \frac{\de\Ga(\Phi,K)}{\de\Phi}\varphi
+ K{\hat s}(\Phi+\varphi)\Big]\Big\}\,.
\eeq
From (\ref{Gexteq}) in tree approximation it follows
\beq
\Gamma^{(0)}(\Phi,K)=S_{FP}(\Phi)+K{\hat s}\Phi.
\eeq

Now we are in a position to write down the Slavnov-Taylor
identity in terms of $\Gamma$,
\beq
\label{WIGd}
\frac{\delta \Gamma}{\delta\Phi^A}
\,\frac{\delta \Gamma}{\delta K_A}\,\equiv\, 0\,.
\eeq
The effective action $\Gamma$  is the main object of study in
quantum theory of Yang-Mills fields, which contains all information
about Green functions of the theory. By construction, $\Gamma$
depends on gauge but this dependence has a very special form. Let
us investigate this dependence. Consider infinitesimal variation
of the gauge function $\chi^a\rightarrow \chi^a+\delta\chi^a$ in
the generating functional of the Green functions,
\ ${\cal Z}={\cal Z}( J,K)$,
\beq
\label{varZ11}
\delta{\cal Z} &=&
\frac{i}{\hbar}\int {\cal D} \Phi\,\Big(
{\bar C}^a\,\frac{\de\,\de\chi^a}{\de A^c_\mu}\,D^{cb}_\mu C^b
+ B^a\de\chi^a\Big)\,
\exp \Big\{\frac{i}{\hbar}\,\big[
S_{FP}({\Phi})+ J\Phi + K{\hat s}\Phi\big])\Big\}
\\
\nonumber
&=&
\frac{i}{\hbar}\int {\cal D}\Phi\,
\bigg[{\bar C}^a\frac{\de\,\de\chi^a}{\de A^{b\mu}}\,
\frac{\de (K{\hat s}\Phi)}{\de K^b_{\mu}}
\,+\, \de\chi^a\,\frac{\de (K{\hat s}\Phi)}{\de L^a}\bigg]
\exp \Big\{\frac{i}{\hbar}
\big[S_{FP}({\Phi}) + J\Phi + K{\hat s}\Phi\big] \Big\}\,.
\eeq
Introducing the functional
\beq
\delta\psi=\delta\psi(\Phi)={\bar C}^a\delta\chi^a, \qquad \varepsilon(\delta\psi)=1
\eeq
we can rewrite (\ref{varZ11}) in the form
\beq
\label{varZ13}
\delta{\cal Z}=\frac{i}{\hbar}\int {\cal D}\Phi
\Big[\frac{\delta\delta\psi}{\delta A^{a\mu}}\frac{\delta (K{\hat
s}\Phi)}{\delta K^a_{\mu}}
+\frac{\de\,\de\psi}{\de{\bar C}^a}\,
\frac{\de(K{\hat s}\Phi)}{\de L^a}\Big]
\exp \Big\{\frac{i}{\hbar}\big[ S_{FP}({\Phi}) + J\Phi
+ K{\hat s}\Phi\big]\Big\}\,.
\eeq
Taking into account that
\beq
\frac{\de\,\de\psi}{\delta C^a}=0
\,,\qquad
\frac{\delta (K{\hat s}\Phi)}{\delta N^a}=0\,,
\eeq
we can present the gauge dependence of generalized generating functional
${\cal Z}$ (\ref{varZ13}) by the following relation
\beq
\label{varZ14}
\delta{\cal Z}\,=\,
\frac{i}{\hbar} \int {\cal D}\Phi
\,\,\frac{\de\,\delta\psi}{\de\Phi^A}
\,\frac{\de (K{\hat s}\Phi)}{\de K_A}\,\,
\exp \Big\{ \frac{i}{\hbar} \big[
S_{FP}({\Phi}) + J\Phi + K{\hat s}\Phi\big] \Big\}\,.
\eeq
It is assumed that the functional integral of total variational
derivative is zero. In this way we arrive at the relation
\beq
\label{EQU}
\int {\cal D}\Phi\,\,\frac{\de}{\de\Phi^A}
\bigg[\de\psi\,\frac{\de (K {\hat s}\Phi)}{\delta K_A}
\exp \Big\{ \frac{i}{\hbar}\,\big[S_{FP}({\Phi}) + J\Phi
+ K{\hat s}\Phi\big]\Big\}\bigg]\,=\,0\,.
\eeq
After a small algebra it can be presented in the form
\beq
\nonumber
&& \int {\cal D}\Phi\,\,
\frac{\de\,\de\psi}{\delta\Phi^A}
\,\frac{\de (K{\hat s}\Phi)}{\delta K_A}\,
\exp \Big\{ \frac{i}{\hbar}\big[ S_{FP}({\Phi}) + J\Phi
+ K{\hat s}\Phi\big]\Big\}
\\
\nonumber
&& \quad = \, -\, \frac{i}{\hbar} \int {\cal D}\Phi\,\de\psi
\Big[ J_A\frac{\de (K{\hat s}\Phi)}{\de K_A}
+ \frac{\de S_{FP}}{\de\Phi^A}
\,\frac{\de (K{\hat s}\Phi)}{\de K_A}
+ \frac{\de (K{\hat s}\Phi)}{\de \Phi^A}
\,\frac{\de (K{\hat s}\Phi)}{\de K_A}
\\
\label{EQU1}
&& \quad + \,\frac{\de^2 (K{\hat s}\Phi)}{\de K_A\de \Phi^A}\Big]
\,\exp \Big\{ \frac{i}{\hbar}\,\big[ S_{FP}({\Phi})
+ J\Phi + K{\hat s}\Phi\big] \Big\}\,.
\eeq
By means of the BRST symmetry of the Faddeev-Popov action
\beq
\frac{\de S_{FP}}{\de\Phi^A}\,\frac{\de (K{\hat s}\Phi)}{\delta K_A}
\,=\,\frac{\de S_{FP}}{\delta\Phi^A}{\hat s}\Phi^A\,=\,0\,,
\end{eqnarray}
using the nilpotency of the BRST transformations
\beq
\frac{\delta (K{\hat
s}\Phi)}{\delta \Phi^A}\frac{\delta (K{\hat
s}\Phi)}{\delta K_A}=K_B{\hat s}\delta^B_A{\hat s}\Phi^A=K_A{\hat s}^2\Phi^A=0
\eeq
and the equality
\beq
\frac{\de^2 (K{\hat s}\Phi)}{\de K_A\,\de \Phi^A}
\,=\,\frac{\de}{\de K_A}
\Big[\frac{\delta(K{\hat s}\Phi)}{\de\Phi^A}\Big]
\,=\,
\frac{\delta}{\delta K_A}\,K_A{\hat s} \cdot 1 \,=\, 0\,,
\eeq
one can reduce the relation (\ref{EQU1}) to the form
\beq
\nonumber
&& \int {\cal D}\Phi
\,\,\frac{\delta\de\psi}{\de\Phi^A}
\,\frac{\de (K{\hat s}\Phi)}{\de K_A}\,\,
\exp \Big\{ \frac{i}{\hbar} \big[S_{FP}({\Phi}) + J\Phi
+ K{\hat s}\Phi\big] \Big\}
\\
\label{EQU2}
&=& -\,\frac{i}{\hbar}\int {\cal D}\Phi\,\de\psi(\Phi)\,
J_A\,\frac{\de (K{\hat s}\Phi)}{\de K_A}
\exp \Big\{ \frac{i}{\hbar} \big[
S_{FP}({\Phi}) + J\Phi + K{\hat s}\Phi\big] \Big\}\,.
\eeq
With the help of (\ref{EQU2}), the gauge dependence of
generalized generating functional \ ${\cal Z}$ \ (\ref{varZ14})
can be rewritten in the form
\beq
\label{varZ15}
\delta{\cal Z} &=& -\,
\Big(\frac{i}{\hbar}\Big)^2 \int {\cal D}\Phi
\,\de\psi(\Phi)\, J_A\,\frac{\delta (K{\hat s}\Phi)}{\delta K_A}
\exp \Big\{\frac{i}{\hbar}\big[ S_{FP}({\Phi}) + J\Phi
+ K{\hat s}\Phi\big] \Big\}
\eeq
and furthermore as
\beq
\label{varZ16}
\delta{\cal Z}(J,K) &=&
\frac{i}{\hbar}\,J_A\,\frac{\delta}{\delta K_A}\,
\delta\psi\Big(\frac{\hbar}{i}\frac{\delta}{\delta J}\Big)
{\cal Z}(J,K)\,.
\eeq
This equation can be rewritten also in terms of the generating
functional of connected Green's functions, \ ${\cal W}$, \ as
\beq
\label{varW}
\delta{\cal W}(J,K)
&=&
J_A \Big(\frac{i}{\hbar}\frac{\delta {\cal W}}{\delta K_A}
+\frac{\delta}{\delta K_A}\Big)
\,\delta\psi \,\Big(\frac{\delta{\cal W}}{\delta J}+ \frac{\hbar}{i}\frac{\delta}{\delta J}\Big)\,.
\eeq
Taken together with the Slavnov-Taylor identity for ${\cal W}$
(\ref{WIWJKI}), the identity (\ref{varW}) can be presented as
\beq
\label{varW1}
\de{\cal W}(J,K) &=&
J_A\, \frac{\de}{\de K_A}\, \de\psi\,
\Big(\frac{\delta{\cal W}}{\delta J}+
\frac{\hbar}{i}\frac{\delta}{\delta J}\Big)\,.
\eeq

Finally, after making Legendre transformation, one can arrive at the
equation describing the gauge dependence of the effective action,
\beq
\label{varG}
\de\Ga(\Phi,K) &=& -\,\frac{\de\Gamma(\Phi,K)}{\de\Phi^A}\,
\frac{\de}{\de K_A}\, \de\psi({\hat\Phi})\,,
\eeq
where the notation
\beq
{\hat\Phi}^A &=&
\Phi^A + \im\hbar\,(\Gamma^{''-1})^{AB}\,
\frac{\delta_l}{\delta\Phi^B}
\eeq
has been used. The matrix \ $(\Gamma^{''-1})$ \ is inverse to the
matrix $\Gamma^{''}$, the last has elements
\beq
(\Gamma^{''})_{AB} &=&
\frac{\de_l}{\de\Phi^A}
\Big(\frac{\de\Gamma}{\de\Phi^B}\Big)
\,,\qquad \mbox{i.e.,} \qquad
\big(\Gamma^{''-1}\big)^{AC}\cdot
\big(\Gamma^{''}\big)_{CB}\,=\,\de^A_{\,B}\,.
\eeq
The main meaning of Eq. (\ref{varG}) is that the effective action
does not depend on the choice of gauge function on-shell, which is
defined by the effective equations of motion,
\beq
\frac{\de \Ga(\Phi,K)}{\de\Phi^A} &=& 0
\qquad \Longrightarrow \qquad \de\Ga(\Phi,K)=0\,.
\label{on shell}
\eeq
Of course, the same statement is valid for any physically relevant
quantity, in particular the elements of the $S$-matrix are gauge
independent in the same sense. This relevant feature can not be
underestimated. Only due to the gauge independence one can
interpret a calculated physical quantity as being independent
on the method of calculation and, finally, consider the result
being well defined. Let us stress that the relations (\ref{varG})
and (\ref{on shell}) are not related to some approximation, such
as, for instance, certain order of the loop expansion. Much on the
contrary, those are very general non-perturbative relations, which
must be provided in a well-defined quantum theory. The general
property of the on-shell gauge independence can be, therefore,
used as a natural test for a few method of deriving quantum
corrections, in both perturbative and non-perturbative approaches.

\section{Functional RG approach for Yang-Mills theories}

\noindent
In this section we are going to briefly present an approach to the
calculation  of effective action, \ $\Gamma$, \ proposed in paper
\cite{Wett-1,Wett-Reu-1,Wett-Reu-2} (for a review of the method
see \cite{Giess} and references therein), based on the concept
of functional renormalization group (FRG). The main idea of the
FRW is to use instead of $\Gamma$ an effective average action,
\ $\Gamma_k$, \ with a momentum-shell parameter $k$, such that
\beq
\lim_{k\rightarrow 0}\Gamma_k=\Gamma\,.
\eeq
For the Yang-Mills theories it was suggested to modify the
Faddeev-Popov action with the help of the specially designed
regulator action $S_k$
\beq
\label{AcSk}
S_k(A,C,{\bar C}) = \int d^{D}x
\Big\{
\frac{1}{2}A^{a\mu}(x)(R_{k,A})^{ab}_{\mu\nu}(x)A^{b\nu}(x)
+ {\bar C}^a(x)(R_{k,gh})^{ab}(x)C^b(x)\Big\}.
\eeq
In what follows we will use the condensed notations,
\beq
\label{Sk}
S_k(A,C,{\bar C}) &=&
\frac{1}{2}\,A^{a\mu}(R_{k,A})^{ab}_{\mu\nu}A^{b\nu}
\,+\, {\bar C}^a(R_{k,gh})^{ab}C^b\,,
\eeq
where regulator functions $R_{k,A}$ and $R_{k,gh}$ do not depend
on the fields and obey the properties
\beq
\nonumber
\lim_{k\rightarrow 0}(R_{k,A})^{ab}_{\mu\nu}\,=\,0
\,,\qquad
\lim_{k\rightarrow 0} (R_{k,gh})^{ab} \,=\,0\,.
\eeq
It is assumed that the regulator functions model the non-perturbative
contributions to the self-energy part of the diagrams, such that the
dependence on the parameter $k$ enables one to get some relevant
information about the scale dependence of the theory beyond the
loop expansion \cite{Polch}. The application of the FRG method lead
to many interesting achievements in many areas of Quantum Field
Theory, Statistical Mechanics and related areas (see, e.g., the
recent reviews \cite{FRG1,FRG2,FRG3,FRG4,FRG5,FRG6} and references
therein).

Our immediate purpose is to check the consistency of the FRW method
based on the introduction of (\ref{Sk}), by exploring the gauge
dependence of the effective average action, including the
on-shell (in)dependence of this special version of effective
action. As in the previous section, we shall restrict consideration
by the pure Yang-Mills theory.

As a starting point, one has to note that the on-shell gauge
independence which we have demonstrated in the previous section,
is essentially dependent on the BRST invariance of the Faddeev-Popov
action (\ref{FPact}). Therefore, the first issue to check in the
new FRG formulation is whether the BRST invariance holds in the
presence of regulator functions. It is an easy exercise to verify
that this is not the case, namely, the action (\ref{AcSk}) is not
invariant under BRST transformations,
\beq
\delta_B \,S_k(A,C,{\bar C}) \,\neq\, 0\,.
\label{no}
\eeq
Let us note that in the limit $k\rightarrow 0$ the BRST
invariance gets restored, but our interest is to use the
concept of effective average action along the FRG trajectory
and not only in its final point. In this case the output
(\ref{no}) should be seen as a warning signal, requesting a
careful investigation of the issue of gauge dependence. The
next step is to see whether this fact leads or not to the
on-shell gauge dependence in this case. This task is much
more complicated than the one described in section 2, hence
we divide it between this and the next two sections.

The generating functional of Green's functions, ${\cal Z}_k$,
is constructed in the form of the functional integral
\cite{Wett-Reu-2}
\beq
\label{Zk}
{\cal Z}_k(J,K) &=&
\int {\cal D}\Phi
\exp \Big\{\frac{i}{\hbar}\big[
S_{FP}(\Phi)+S_k(\Phi)+J\Phi+K{\hat s}\Phi\big]\Big\}\,,
\eeq
where, for the sake of uniformity, we used notation $S_k(\Phi)$
instead $S_k(A,C,{\bar C})$, despite $S_k$ does not depend on
fields $B^a$. \ In the limit $k\rightarrow 0$ this functional
coincides with the generating functional (\ref{Zext}).

The Slavnov-Taylor identity can be seen as a consequence of
BRST invariance of the Faddeev-Popov action. In the case of
the functional (\ref{Zk}) this identity can be presented in
the form
\beq
\label{STSk}
J_A\frac{\delta {\cal Z}_k}{\delta K_A} \,-\,
i\hbar\Big\{
(R_{k,A})^{ab}_{\mu\nu}
\,\frac{\de^2{\cal Z}_k}{\de j^b_{\nu}\de K^a_{\mu}}
\,+\,
(R_{k,gh})^{ab}\,\frac{\de^2 {\cal Z}_k}{\de\eta^a\de {\bar L}^b}
\,-\,
(R_{k,gh})^{ab}\,\frac{\de^2 {\cal Z}_k}{\de {\bar\eta}^b\de L^a}\Big\}\,\equiv\,0\,.
\eeq
In the limit \ $k\rightarrow 0$ \ the last identity reduces to
(\ref{WIZJKI}). In terms of generating functional of connected
Green's functions, \ ${\cal W}_k={\cal W}_k(J,K)$, \ the Slavnov-Taylor
identity can be written in the form
\beq
\nonumber
&& J_A\frac{\delta {\cal W}_k}{\delta K_A}
\,+\,
\Big\{
(R_{k,A})^{ab}_{\mu\nu}\,\frac{\de {\cal W}_k}{\de j^b_{\nu}}
\,\frac{\de {\cal W}_k}{\de K^a_{\mu}}
\,+\,
(R_{k,gh})^{ab}\,\frac{\de {\cal W}_k}{\de \eta^a}
\,\frac{\de {\cal W}_k}{\de {\bar L}^b}
\,-\, (R_{k,gh})^{ab}\,\frac{\de {\cal W}_k}{\de {\bar\eta}^b}
\frac{\de {\cal W}_k}{\de L^a}\Big\}
\\
\label{WIWkJ}
&-& i\hbar\Big\{
(R_{k,A})^{ab}_{\mu\nu} \,
\frac{\de^2 {\cal W}_k}{\de j^b_{\nu}\delta K^a_{\mu}}
\,+\,
(R_{k,gh})^{ab}\frac{\de^2 {\cal W}_k}{\de \eta^a\delta {\bar L}^b}
\,-\,
(R_{k,gh})^{ab}\,\frac{\de^2 {\cal W}_k}{\de {\bar\eta}^b\de L^a}\Big\}
\,\equiv\, 0\,.
\eeq

Finally, we introduce the generating functional of vertex functions
in the presence of regulators (the effective average action),
$\Gamma_k=\Gamma_k(\Phi,K)$, as
\beq
\label{LtrW}
\Gamma_k(\Phi,K) &=& {\cal W}_k(J,K)-J\Phi
\,,\qquad
\Phi^A=\frac{\delta {\cal W}_k}{\delta J_A}\,.
\eeq
In the last expression the source $J_A$ is regarded as a
function of the mean field $\Phi^A$. Then
\beq
\label{LtrG}
\frac{\de\Gamma_k}{\de \Phi^A} = - J_A
\,,\qquad
\frac{\de\Gamma_k}{\de K_A} = \frac{\de {\cal W}_k}{\de K_A}\,.
\eeq
The functional \ $\Gamma_k$ \ satisfies the functional
integro-differential equation
\beq
&& \exp \Big\{ \frac{i}{\hbar}\,\Ga_k(\Phi,K) \Big\}
\label{GkEq}
\\
\nonumber
&=&
\int {\cal D}\varphi \,\exp \Big\{\frac{i}{\hbar}
\Big[S_{FP}({\Phi+\varphi})+S_k(\Phi+\varphi)
\,+\,K{\hat s}(\Phi+\varphi)\,-\,\frac{\de \Ga_k(\Phi,K)}{\de\Phi}\,\varphi\Big]\Big\}\,.
\eeq
The tree-level (zero-loop) approximation of (\ref{GkEq})
corresponds to
\beq
\Gamma^{(0)}_k(\Phi,K)\,=\,
S_{FP}({\Phi}) \,+\, S_k(\Phi) \,+\, K{\hat s}\Phi\,.
\eeq
It proves useful to introduce another version of effective action,
which does not depend on external sources \ $K_A$, such that the
modified version of equation (\ref{GkEq}) gets simplified. For this
end we define the functional \ ${\bar \Ga}_k$ \ according to
\beq
{\bar \Gamma}_k &=& \Gamma_k - K{\hat s}\Phi\,.
\label{new EA}
\eeq
One can immediately find that \ ${\bar \Ga}_k$ \ does not
depend on \ $K$ \ and satisfies the equation
\beq
\label{GkbarEq}
\exp \Big\{
\frac{i}{\hbar}\,{\bar\Ga}_k(\Phi)\Big\}
&=& \int {\cal D}\varphi
\,\exp \Big\{ \frac{i}{\hbar} \Big[
S_{FP}({\Phi+\ph})+S_k(\Phi+\ph)
- \frac{\de {\bar\Ga}_k(\Phi)}{\de\Phi}\,\varphi\Big]\Big\}\,.
\eeq
The derivation of the Slavnov-Taylor identity from the the BRST
symmetry follows the same steps which we described in details
in the previous section, so we can present just the final form
of this identity in terms of \ $\Gamma_k=\Gamma_k(\Phi,K)$.
\ The result reads
\beq
\nonumber
\frac{\delta\Gamma_k}{\delta\Phi^A}\,
\frac{\delta \Gamma_k}{\delta K_A}
&-& \Big\{(R_{k,A})^{ab}_{\mu\nu}\,A^{b\nu}\,
\frac{\de\Ga_k}{\de K^a_{\mu}}
\,+\, (R_{k,gh})^{ab}\,{\bar C}^a\,\frac{\de \Ga_k}{\de {\bar L}^b}
\,-\, (R_{k,gh})^{ab}\,C^b\,\frac{\de \Ga_k}{\de L^a}\Big\}
\\
\nonumber
&-&
i\hbar\Big\{
(R_{k,A})^{ab}_{\mu\nu}\,\big(\Gamma^{''-1}\big)^{b\nu\,A}
\,\frac{\de^2_{l} \Ga_k}{\de \Phi^A\,\de K^a_{\mu}}
\,+\, (R_{k,gh})^{ab}\,\big(\Gamma^{''-1}_k\big)^{aA}\,
\frac{\de^2_{l}\Ga_k}{\de \Phi^A\,\de {\bar L}^b}
\\
\label{STGk}
&-&
(R_{k,gh})^{ab}\,\big(\Gamma^{''-1}_k\big)^{{\bar b}A}\,
\frac{\de^2_{l}\Ga_k}{\de\Phi^A\,\delta L^a}
\Big\}\,\equiv\, 0\,.
\eeq
The matrix $(\Gamma^{''-1}_k)$ is inverse to the matrix
$\Gamma^{''}_k$ with elements
\beq
\label{invGk}
(\Gamma^{''}_k)_{AB}\=\frac{\delta_l}{\delta\Phi^A}
\Big(\frac{\delta\Gamma_k}{\delta\Phi^B}\Big)\,,
\qquad \mbox{i.e.,}\qquad
\big(\Gamma^{''-1}_k\big)^{AC}\,
\big(\Gamma^{''}_k\big)_{CB}\,=\,\de^A_{\ B}\,.
\eeq
In the last expressions we have used the following notation for indices:
\ $A=\big( (a\mu), a, {\bar a}, {\tilde a}\big)$, corresponding
to the fields \ $\Phi^A=\big(A^{a\mu}, C^a, {\bar C}^a, B^a\big)$.
In the zero-loop approximation,
\ $\Gamma_k(\Phi,K)=\Gamma^{(0)}_k(\Phi,K)$, then the identity
(\ref{STGk}) reduces to the Zinn-Justin equation \cite{Z-J} for
the action \ $S_{ext}(\Phi,K)=S_{FP}(\Phi)+K{\hat s}\Phi$, \ namely to
\beq
\nonumber
\frac{\de S_{ext}}{\de\Phi^A}\,\frac{\de S_{ext}}{\de K_A}\,=\,0\,,
\eeq
that corresponds to the BRST symmetry of the Faddeev-Popov action.
Note that the Slavnov-Taylor identity in momentum space for the FRG
approach was previously considered in the work \cite{Ell}.

The FRG flow equation written for the generating functional
\ ${\cal W}_k={\cal W}_k(J,K)$ \ can be written in detailed form as
follows:
\beq
\nonumber
\partial_{t}{\cal W}_k
&=&
\frac{1}{2}\,\pa_t (R_{k,A})^{ab}_{\mu\nu}\,
\frac{\de {\cal W}_k}{\de j^a_{\mu}}
\,\frac{\de {\cal W}_k}{\de j^b_{\nu}}
\,+\,
\pa_t (R_{k,gh})^{ab}\,
\frac{\de {\cal W}_k}{\de \eta^a}\,
\frac{\de {\cal W}_k}{\de {\bar\eta}^b}
\\
&-& i\hbar\Big\{
\frac{1}{2}\,\pa_t (R_{k,A})^{ab}_{\mu\nu}\,
\frac{\de^2 {\cal W}_k}{\de j^a_{\mu}\,\de j^b_{\nu}}
\,+\, \pa_t (R_{k,gh})^{ab}
\,\frac{\de^2 {\cal W}_k}{\de \eta^a\,\de {\bar\eta}^b}\Big\}\,,
\eeq
where we used a standard notation
$$
\pa_t = k\,\frac{d}{dk}
$$
and took into account that the dependence on $k$ comes only from
the corresponding dependence of the regulator functions (\ref{AcSk}).

Consider the FRG flow equation for the effective average action,
\ $\Ga_k$. From the definition of $\Phi$ in (\ref{LtrW}) one can
see that it is dependent on the parameter \ $k$, that means
\ $\pa_t\Phi^A \neq 0$.
Therefore, one has to be very careful in calculations and take
into account all ways of $k$-dependence. We have
\beq
\partial_t\Gamma_k\Big |_{\Phi,K}+
\frac{\delta\Gamma_k}{\delta\Phi^A}\Big |_{k,K}\partial_t\Phi^A=
\partial_t{\cal W}_k\Big |_{J,K}-J_A\partial_t\Phi^A\Big |_{J,K}\,,
\eeq
where the index $\big|_X$ after partial or variational derivative
means that the quantity $X$ is kept constant. Due to the properties
of the Legendre transformations (\ref{LtrG}), we obtain
\beq
\partial_t\Gamma_k\Big |_{\Phi,K} \,=\,
\partial_t{\cal W}_k\Big |_{J,K}\,.
\eeq

As far as the FRG parameter $k$ is not physical, we can not expect
that it will emerge in the final physical output of the theory.
Therefore, summing up all types of $k$-dependence we always get
zero. The application of the renormalization group method implies
that we take into account only part of $k$-dependence, e.g., find
how the effective action depends on $k$ and then trade it for
some physical parameter corresponding to the problem of our
interest (see detailed discussion of this issue in \cite{DCCR}).
Therefore, in the FRG flow equation for $\Gamma_k$, only the
explicit dependence on $k$ should be taken into account, so we get
\beq
\label{mEqRG}
\pa_t\Gamma_k \,=\, \pa_tS_k
\,+\, i\hbar\,\,\Big\{
\frac{1}{2}\,\pa_t(R_{k,A})^{ab}_{\mu\nu}\,
\big(\Gamma^{''-1}_k\big)^{(a\mu)(b\nu)}
\,+\,\pa_t(R_{k,gh})^{ab}\,\big(\Ga^{''-1}_k\big)^{ab}\Big\}\,.
\eeq
Usually the functional RG approach is formulated in terms of the
functional which does not depend on sources $K$. Since the equation
(\ref{mEqRG}) does not contain derivatives with respect to $K$, we
can just put $K_A=0$ and arrive at
\beq
\label{mEqRG1}
\pa_t{\bar \Gamma}_k &=&
\pa_t S_k
\,+\,
i\hbar\,\Big\{
\frac{1}{2}\,\pa_t(R_{k,A})^{ab}_{\mu\nu}\,
\big(
{\bar\Gamma}^{''-1}_k\big)^{(a\mu)(b\nu)}
\,+\,
\pa_t(R_{k,gh})^{ab}\,
\big({\bar\Gamma}^{''-1}_k\big)^{ab}\Big\}\,,
\eeq
where
$$
{\bar \Gamma}_k={\bar \Gamma}_k(\Phi)=\Gamma_k\big(\Phi,K=0\big)\,.
$$
${\bar \Gamma}_k(\Phi)$ satisfies equation (\ref{GkbarEq}). In the
condensed notations we can write down the equation (\ref{mEqRG1})
in the form
\beq
\label{mEqRG1-n}
\pa_t{\bar \Gamma}_k &=&
\pa_t S_k
\,+\,
i\hbar\,\Big\{
\frac{1}{2}\,
\Tr \big[\pa_t(R_{k,A})\big({\bar\Gamma}^{''-1}_k\big)\big]_A
\,-\,\Tr \big[\pa_t(R_{k,gh})\big({\bar\Gamma}^{''-1}_k\big)
\big]_C\Big\}\,,
\eeq
where we took into account the anticommuting nature of the
ghost fields $C^A$ and defined
\beq
\Tr \Big[\pa_t(R_{k,gh}) \big({\bar\Gamma}^{''-1}_k\big)\Big]_C
&=& -\,\pa_t(R_{k,gh})^{ab} \big({\bar\Gamma}^{''-1}_k\big)^{ab}\,,
\nonumber
\\
\Tr \Big[\pa_t(R_{k,A}) \big({\bar\Ga}^{''-1}_k\big)\Big]_A
&=& \pa_t(R_{k,A})^{ab}_{\mu\nu}
\big(\Ga^{''-1}_k\big)^{(a\mu)(b\nu)}\,.
\eeq

In the tree-level approximation we have \ ${\bar\Gamma}_k=S_{FP}+S_k$
\ and the equation (\ref{mEqRG1-n}) is satisfied because the
Faddeev-Popov action does not depend on \ $k$. The conventional
presentation of the FRG flow equation is in terms of the functional
${\it{\Gamma}_k}={\bar\Gamma}_k-S_k$, which satisfies the equation
\beq
\pa_t{\it{\Gamma}}_k
\,=\,
i\hbar\,\Big\{
\frac{1}{2}\,\Tr \Big[\pa_tR_{k,A}\,\big({\it{\Gamma}}^{''}_k
+ R_{k,A}\big)^{-1}\Big]_A
\,-\, \Tr \Big[\pa_t (R_{k,gh})\,
\big({\it {\Gamma}}^{''}_k + R_{k,gh}\big)^{-1}\Big]_C
\Big\}\,.
\eeq
The last representation uses symbolic notations of inverse
matrices \ $({\it{\Gamma}}^{''}_k+R_{k,A})^{-1}$
\ and \ $({\it {\Gamma}}^{''}_k+R_{k,gh})^{-1}$,
which are in fact components of the following inverse matrix
\beq
\big({\it \Gamma}_k+S_k\big)^{''}_{AB}
&=& \frac{\delta_l}{\delta\Phi^A}\,
\Big[\frac{\de ({\it \Ga}_k + S_k)}{\de\Phi^B}\Big]
\eeq
in the sectors of vector and ghost fields, respectively.

\section{Vacuum functional in the FRG for Yang-Mills field}

\noindent
Let us explore the vacuum functional in the FRG approach. We
introduce the notation $Z_{k,\chi}$ for the vacuum functional,
corresponding to the given gauge function $\chi^a$ in the
Faddeev-Popov action
\beq
Z_{k,\chi}={\cal Z}_k(0,0)=\int {\cal D}\Phi\;
{\rm exp}\Big\{\frac{i}{\hbar}\big(S_{FP}+S_k\big)\Big\}\,.
\eeq
By construction, the regulator functions in the FRG approach do
not depend on gauge  $\chi^a$ and therefore the action $S_k$ is
gauge independent. Let us consider an infinitesimal variation of
gauge $\chi \rightarrow \chi+\delta\chi$ and construct the vacuum
functional corresponding to this gauge. We have
\beq
\label{Zkvar}
Z_{k,\chi+\delta\chi} &=&
\int {\cal D}\Phi\;
\exp \Big\{ \frac{i}{\hbar}\Big(S_{FP}+S_k+
{\bar C}^a\,\frac{\de\,\de\chi^a}{\delta A^c_{\mu}}\,
D^{cb}_{\mu}C^b + \de\chi^aB^a\Big)\Big\}\,.
\eeq
In the functional integral (\ref{Zkvar}) we make a change
of variables in the form of the BRST transformations
(\ref{BRSTtr}), but trading the constant Grassmann-odd
parameter $\theta$ to a functional $\Lambda=\Lambda(\Phi)$.
The Faddeev-Popov action, $S_{FP}$, is invariant under such
change of variables but $S_k$ is not invariant, with the
variation given by
\beq
\label{Skvar}
\delta S_k \,=\,
A^{a\mu}(R_{k,A})^{ab}_{\mu\nu}\,D^{\nu bc}C^c\Lambda
\,+\,\frac{1}{2}\,{\bar C}^a(R_{k,gh})^{ab}\,f^{bcd}\,C^c C^d
\Lambda
\,-\,B^a(R_{k,gh})^{ab}\,C^b\Lambda.
\eeq
The contributions which come from  the measure of functional
integral have the form\footnote{Compare to the considerations
leading to Eq. (\ref{FPVac1}).}
\beq
\label{Mvar}
i\hbar \Big(\,\frac{\de\La}{\de A^a_\mu}\,D^{ab}_{\mu}\,C^b
\,-\, \frac{1}{2}\,f^{abc}\,C^b C^c\,\frac{\de\La}{\de C^a}
\,+\,\frac{\de\La}{\de {\bar C}^a}\,B^a\Big)\,.
\eeq

Let us discuss the possibility to compensate the variation
(\ref{Skvar}) by choosing a special form of the functional
$\La$. The transformed value of the generating functional
can be obtained by combining the formulas
(\ref{Zkvar}), (\ref{Skvar}) and (\ref{Mvar}),
\beq
Z_{k,\chi+\delta\chi} &=&
\int {\cal D}\Phi\;
\exp \Big\{ \frac{i}{\hbar}\Big[S_{FP}(\Phi) + S_k
\,+\, {\bar C}^a\,\frac{\de\,\de\chi^a}{\de A^c_{\mu}}\,
D^{cb}_{\mu}C^b
\,-\, B^{a}\,\big(R_{k,gh}\big)^{ab}\,C^b\La
\nonumber
\\
&+& \frac12\,{\bar C}^a\,\big(R_{k,gh}\big)^{ab}
\,f^{bcd}\,C^cC^d\,\La
\,+\,
A^{a\mu}\,\big(R_{k,A}\big)^{ab}_{\mu\nu}\,D^{\nu\,bc}\,C^c\La
\,+\, \de\chi^aB^a
\nonumber
\\
&-& \frac{i\hbar}{2}\,f^{abc}\,C^bC^c\,\frac{\de\La}{\de C^a}
\,+\,
i\hbar\,\frac{\de\La}{\de A^a_\mu}\,D^{ab}_\mu\,C^b
\,+\,
i\hbar\,\frac{\de\La}{\de {\bar C}^a}\,B^a\Big]\Big\}\,.
\label{transZk}
\eeq
In order to provide compensation of all new terms in
$Z_{k,\chi+\delta\chi}$, such that it becomes equal to
$Z_{k,\chi}$, one has to satisfy the following equations:
\beq
&&
i\hbar\,\frac{\de\La}{\de A^a_\mu}
\,-\,
A^{b\nu}\,\big(R_{k,A}\big)^{ba}_{\nu\mu}\,\La
\,+\,{\bar C}^b\,\frac{\de\,\de\chi^b}{\de A^a_{\mu}}
\,=\, 0\,,
\nonumber
\\
&&
i\hbar\,\frac{\de\La}{\de {\bar C}^a}
\,+\,\delta\chi^a
\,-\,\big(R_{k,gh}\big)^{ab}\,C^b\La \,=\, 0\,,
\\
&&
i\hbar\,\frac{\de\La}{\de C^a}
\,+\,
\frac12\,{\bar C}^b\,\big(R_{k,gh}\big)^{ba}\La \,=\, 0\,.
\label{eqsZk}
\eeq
It is easy to see that the solution of the last equation,
\beq
\frac{\de\La}{\de C^a}
 &=&\frac{1}{2\,i\hbar}\,{\bar C}^b\,\big(R_{k,gh}\big)^{ba}\La\,,
\label{lasteqsZk}
\eeq
has the form
\beq \La &=&
\frac{1}{2\,i\hbar}\,{\bar C}^a \,\big(R_{k,gh}\big)^{ab}\,C^b \,+\,
\La_1({\bar C},B,A)\,. \label{soleqsZk}
\eeq
At the same time, the
first of the equations (\ref{eqsZk}) can be cast into the form
\beq
i\hbar\,\frac{\de\La_1}{\de A^a_\mu} \,+\,{\bar
C}^b\,\frac{\de\,\de\chi^b}{\de A^a_{\mu}}
\,-\,A^{b\nu}\,\big(R_{k,A}\big)^{ba}_{\nu\mu}\,\La_1
\,-\,\frac{1}{2\,i\hbar}\,
A^{b\nu}\,\big(R_{k,A}\big)^{ba}_{\nu\mu}\,{\bar C}^d\,
\big(R_{k,gh}\big)^{dc}\,C^c = 0 \label{1st-eqsZk}
\eeq and its
solution should be dependent on the field $C^a$ as well. One can
conclude that this contradicts the (\ref{soleqsZk}) and, therefore,
it is not possible to compensate the variation (\ref{Skvar}) by
choosing a special form of the functional $\La$.

For instance, if we choose $\La$ in a natural way, \beq \La =
\frac{i}{\hbar}\, {\bar C}^a\,\delta\chi^a\,, \nonumber \eeq then
\beq \label{Zkvar1} Z_{k,\chi+\delta\chi}\,=\, \int {\cal D}\Phi\,
\exp \Big\{\frac{i}{\hbar}\big(S_{FP} + S_k + \de S_k \big)\Big\}\,.
\eeq Then, for any value $k\neq 0$, one has \beq
Z_{k,\chi+\delta\chi}\neq Z_{k,\chi}. \eeq Therefore the vacuum
functional $Z_{k,\chi}$ (and therefore $S$-matrix) depends on gauge.

One might think that situation with gauge dependence within the
FRG approach can be improved if we propose gauge dependence of
regulators \ $(R_{k,A})^{ab}_{\mu\nu}$ \ and \ $(R_{k,gh})^{ab}$,
such that the gauge variation of $S_k$ gives additional contributions
\beq
\frac{1}{2}\,A^{a\mu}\de(R_{k,A})^{ab}_{\mu\nu}\,A^{b\nu}
\,+\, {\bar C}^a\,\de (R_{k,gh})^{ab}\,C^b\,.
\label{desespe}
\eeq
Unfortunately, one can easily verify that the terms appearing in the
exponent inside the functional integral cannot be compensated
by choosing the functional $\Lambda$. Therefore, the possible
generalization (\ref{desespe}) can not solve the problem of
gauge dependence of vacuum functional within the FRW approach.

\section{Gauge dependence of $\Gamma_k$}

\noindent
Let us explore the gauge dependence of the generating
functionals \ ${\cal Z}_k$, ${\cal W}_k$ and $\Gamma_k$ for
Yang-Mills theory in the framework of the FRG approach. We shall
follow the general methods from the papers \cite{VLT-82,L}. The
derivation of this dependence is based on a variation of the
gauge-fixing function, $\chi^a\rightarrow\chi^a+\delta\chi^a$, which
leads to the variation of the Faddeev-Popov action $S_{FP}$
(\ref{FPact}) and consequently of the generating functional
${\cal Z}_k={\cal Z}_k(J,K)$ (\ref{Zk}). Introducing the functional
$\delta\psi={\bar C}^a\delta\chi^a$, the gauge dependence can be
presented in the form
\beq
\label{varZk}
\delta{\cal Z}_k &=&
\frac{i}{\hbar}\int {\cal D}\Phi
\,\,\frac{\de\,\de\psi}{\delta\Phi^A}\,\, \frac{\delta (K{\hat
s}\Phi)}{\delta K_A}\, \exp \Big\{\frac{i}{\hbar}\big[S_{FP}({\Phi})
+ S_k + J\Phi + K{\hat s}\Phi\big] \Big\}\,.
\eeq

Let us consider an obvious relation
\beq
\label{EQU-bis}
\int {\cal D} \Phi\,\,\frac{\de}{\de\Phi^A}
\bigg\{
\de\psi\,\frac{\de (K{\hat s}\Phi)}{\de K_A}
\,\exp \Big\{\frac{i}{\hbar}\big[S_{FP}({\Phi})
+ S_k + J\Phi + K {\hat s}\Phi\big]\Big\}\bigg\}\,=\,0\,.
\eeq
It is not difficult to obtain its extended form,
\beq
\label{EQU2-bis}
&&\int {\cal D}\Phi
\,\,\frac{\de\,\de\psi}{\de\Phi^A}\,
\frac{\de (K{\hat s}\Phi)}{\de K_A}\,
\exp \Big\{\frac{i}{\hbar}\big[ S_{FP}({\Phi})+S_k+J\Phi
+K{\hat s}\Phi\big]\Big\}
\\
&& \quad =\,-\,
\frac{i}{\hbar}\int {\cal D}\Phi\,\,
\de\psi(\Phi)\,\Big(J_A+\frac{\de S_k}{\de\Phi^A}\Big)
\,\frac{\de (K{\hat s}\Phi)}{\de K_A}
\exp \Big\{ \frac{i}{\hbar}\big[
S_{FP}({\Phi})+S_k+J\Phi +K{\hat s}\Phi\big]\Big\}\,.
\nonumber
\eeq
Using the last formula, we can rewrite gauge dependence of
the functional ${\cal Z}_k$ in the form
\beq
\nonumber
\delta{\cal Z}_k=-\Big(\frac{i}{\hbar}\Big)^2\int {\cal D}\Phi\;
\delta\psi(\Phi)\Big(
J_A+\frac{\delta S_k}{\delta\Phi^A}\Big)\frac{\delta (K{\hat
s}\Phi)}{\delta K_A}
\exp \Big\{ \frac{i}{\hbar}\big[
S_{FP}({\Phi})+S_k+J\Phi +K{\hat s}\Phi\big]\Big\}\,.
\eeq
Taking into account the explicit structure of $S_k$ (\ref{Sk})
one can get the relation
\beq
\frac{\de S_k}{\de\Phi^A}\,\frac{\de (K{\hat s}\Phi)}{\de K_A}
&=&
A^{a\mu}(R_{k,A})^{ab}_{\mu\nu}\,
\frac{\de (K{\hat s}\Phi)}{\de j^b_{\nu}}
\nonumber
\\
&+&
{\bar C}^a(R_{k,gh})^{ab}\,
\frac{\de (K{\hat s}\Phi)}{\de {\bar L}^b}
\,-\,
C^a(R_{k,gh})^{ba}\,
\frac{\de (K{\hat s}\Phi)}{\de L^b}\,.
\eeq
For the final form of gauge dependence of generating functional
 ${\cal Z}_k={\cal Z}_k( J,K)$ we obtain the relation
\beq
\label{varZkf}
&&\delta{\cal Z}_k=\frac{i}{\hbar}J_A\frac{\delta }{\delta K_A}
\delta\psi\Big(\frac{\hbar}{i}\frac{\delta}{\delta J}\Big)
{\cal Z}_k+
\\
\nonumber
&+&
\Big[(R_{k,A})^{ab}_{\mu\nu}
\frac{\delta^2 }{\delta j^a_{\mu}\delta j^b_{\nu}}
+ (R_{k,gh})^{ab}
\frac{\delta^2}{\delta \eta^a\delta {\bar L}^b}
- (R_{k,gh})^{ba}
\frac{\delta^2}{\delta {\bar\eta}^a\delta L^b}\Big]
\,\de\psi\Big(\frac{\hbar}{i}\frac{\de}{\de J}\Big)\,{\cal Z}_k\,.
\eeq
The last equation can be rewritten in terms of the generating
functional of the connected Green's functions, \
${\cal W}_k={\cal W}_k(J,K)$, to give
\beq
\label{varWk}
\delta{\cal W}_k
&=&
J_A\,\frac{\de}{\de K_A}\,\de\psi\,
\Big(\frac{\de{\cal W}_k}{\de J}
+ \frac{\hbar}{i}\,\frac{\de}{\de J}\Big)
\,-\,
i\hbar
\Big[(R_{k,A})^{ab}_{\mu\nu}\,
\frac{\de^2}{\de j^a_\mu\,\de j^b_\nu}
\\
&+&
(R_{k,gh})^{ab}\,\frac{\de^2}{\de \eta^a\delta {\bar L}^b}
\,-\,
(R_{k,gh})^{ba}
\frac{\delta^2}{\delta {\bar\eta}^a\delta L^b}\Big]
\delta\psi\Big(\frac{\delta{\cal W}_k}{\delta J}+
\frac{\hbar}{i}\frac{\delta}{\delta J}\Big)\,,
\nonumber
\eeq
where the identity (\ref{WIWkJ}) was used.

The next step is to write the corresponding equation for
the effective average action. It proves convenient to
introduce the following notation:
\beq
S_{k;A}(\Phi)=\frac{\delta S_k(\Phi)}{\delta\Phi^A}.
\label{not}
\eeq
Then the equation (\ref{varWk}) reads
\beq
\delta{\cal W}_k
&=&
\Big\{J_A - i\hbar
S_{k;A}\Big(\frac{\de}{\de J}\Big)\Big\}
\,\frac{\de}{\de K_A}
\,\de\psi\Big(\frac{\delta{\cal W}_k}{\delta J}
+ \frac{\hbar}{i}\,\frac{\de}{\de J}\Big)\,.
\eeq
In terms of the effective average action this equation becomes
\beq
\de \Ga_k
&=&
-\,\frac{\de\Ga_k}{\de\Phi^A}\,\frac{\de}{\de K_A}\,
\de\psi\big({\hat \Phi}\big)
\,-\,i\hbar\, S_{k;A}\,\Big (\frac{i}{\hbar}
(\Phi- {\hat\Phi})\Big )\,\frac{\de}{\de K_A}
\,\de\psi\big({\hat \Phi}\big)\,,
\eeq
where ${\hat\Phi^A}$ is defined as
\beq
{\hat\Phi}^A
\,=\,
\Phi^A + \im\hbar\,\big(\Gamma_k^{''-1}\big)^{AB}
\,\frac{\de_l}{\de\Phi^B}\,.
\eeq
The inverse matrix $(\Gamma_k^{''-1})^{AB}$ was introduced
in (\ref{invGk}). We see that if on-shell is defined in the
usual way,
\beq
\frac{\delta\Gamma_k}{\delta\Phi^A} &=& 0\,,
\eeq
then even on-shell the effective average action $\Gamma_k$
depends on gauge, \ $\de \Ga_k\neq 0$. This result
confirms the one of the previous section and shows that the
gauge dependence represents a serious problem for the FRG
approach in the standard conventional formulation.

\section{An alternative approach with composite operators}

\noindent
In this section we are going to suggest an approach in spirit
of the FRG, which is free of the gauge dependence problem for
Yang-Mills theory. This new approach is based on implementing
regulator functions by means of composite fields.

Effective action for composite fields in Quantum Field Theory was
introduced in \cite{CJT}\footnote{Recently, there were interesting
publications discussing the 2PI and $n$PI effective actions, related
to composite fields, to the exact renormalization group
\cite{Paw-2,BPR-2011,Car-2012,IIS-2009} (see also further references
therein). In what follows we develop qualitatively distinct approach
to the FRG, but also using ideas of composite fields.}. Later on,
effective action for composite fields in gauge theories was
introduced and studied in the papers \cite{L1,LO,LOR}. In the case
of gauge theories this effective action depends on gauge. It was
shown that this dependence has a very special form and that there is
a possibility to define a theory with composite fields in such a way
that the effective action of these fields becomes gauge independent
on-shell.

Let us see how the composite fields idea can be used in
the FRG framework for gauge theories. The idea is to use such
a fields to implement regulator functions. Consider the
regulator functions
\beq
\label{L1}
L^1_k(x) &=& \frac{1}{2}\,
A^{a\mu}(x)(R_{k,A})^{ab}_{\mu\nu}(x)A^{b\nu}(x)\,,
\\
\label{L2}
L^2_k(x) &=& {\bar C}^a(x)(R_{k,gh})^{ab}(x)C^b(x)\,.
\eeq
Now we introduce external scalar sources $\Sigma_1(x)$ and
$\Sigma_2(x)$ and construct the generating functional of Green's
functions for Yang-Mills theories with composite fields
\beq
\label{Zcf}
{\cal Z}_k(J,K;\Sigma) &=&
\int {\cal D} \Phi\,
\exp \Big\{\frac{i}{\hbar}\,\big[S_{FP}(\Phi)+J\Phi+K{\hat s}\Phi
+ \Sigma L_k(\Phi)\big]\Big\}\,.
\eeq
where \ $\Sigma L_k=\Sigma_1 L^1_k+\Sigma_2 L^2_k$. \

The generating functional (\ref{Zcf}) may be regarded as a
generalization of the generating functionals discussed previously,
the difference is related to the new sources $\Sigma_1(x)$ and
$\Sigma_2(x)$ and to the corresponding composite fields.
By choosing the sources $\Sigma$ to have zero values,
\ $\Sigma_1(x)=\Sigma_2(x)=0$, the functional (\ref{Zcf})
boils down to the generating functional for Yang-Mills
theories described in Sect. 2. At the same time, when
\ $\Sigma_1(x)=\Sigma_2(x)=1$, the functional
(\ref{Zcf}) corresponds to the generating functional in
the FRG approach introduced in Sect. 3.

One can note that at least one advantage of the proposal
(\ref{Zcf}) is quite evident from the very beginning. If we
define the vacuum functional where all the sources are
switched off, \ $J=K=\Sigma=0$, \ then it coincides with
the vacuum functional for Yang-Mills theory and hence does
not depend on the gauge fixing. Let us now see how this
feature concerns vacuum functional and effective action.

Consider the FRG flow equations in the approach with composite
operators, based on the new generating functional (\ref{Zcf}).
Starting from (\ref{Zcf}) one can repeat the considerations
of Sect. 3 and arrive at the new version of the FRG flow
equation for the generating functional \
${\cal Z}_k={\cal Z}_k(J,K;\Sigma)$,
\beq
\pa_t{\cal Z}_k &=&
\frac{\hbar}{i}\,
\Big\{
\frac{1}{2}\,
\Sigma_1\,\pa_t(R_{k,A})^{ab}_{\mu\nu}\,
\frac{\de^2 {\cal Z}_k}{\de j^a_{\mu}\,\de j^b_{\nu}}
\,+\,
\Sigma_2\,\pa_t(R_{k,gh})^{ab}\,
\frac{\de^2 {\cal Z}_k}{\de \eta^a\,\de {\bar\eta}^b}\Big\}\,.
\label{NflowZ}
\eeq
As usual, we should rewite it in terms of \
${\cal W}_k={\cal W}_k(J,K;\Sigma)$,
\beq
\nonumber
\pa_t{\cal W}_k &=&
\frac{1}{2}\,\Si_1\,\pa_t(R_{k,A})^{ab}_{\mu\nu}\,
\frac{\de {\cal W}_k}{\de j^a_{\mu}}\,
\frac{\de {\cal W}_k}{\de j^b_{\nu}}
\,+\,
\Si_2\,\pa_t(R_{k,gh})^{ab}\,
\frac{\de{\cal W}_k}{\de \eta^a}
\,\frac{\de{\cal W}_k}{\delta {\bar\eta}^b}
\\
&-& i\hbar\,\bigg\{
\frac{1}{2}\, \Si_1\pa_t(R_{k,A})^{ab}_{\mu\nu}\,
\frac{\de^2 {\cal W}_k}{\de j^a_{\mu}\,\de j^b_{\nu}}
\,+\,
\Si_2\,\pa_t(R_{k,gh})^{ab}\,
\frac{\de^2 {\cal W}_k}{\de \eta^a\,\de {\bar\eta}^b}\bigg\}\,.
\label{NflowW}
\eeq

The effective average action with composite fields,
$\Gamma_k=\Gamma_k(\Phi,K;F)$, can be introduced by means of the
following double Legendre transformations (see \cite{CJT} for
the details in case of usual effective action):
\beq
\label{Gcf1} \Ga_k(\Phi,K;F) \,=\, {\cal W}_k(J,K;\Sigma)
\,-\,J_A\Phi^A\,-\, \Si_i\big[L^i_k(\Phi)+\hbar F^i\big]\,,
\eeq
where
\beq \label{Gcf2} \Phi^A=\frac{\delta{\cal W}_k}{\delta J_A}
\,,\qquad\hbar F^i=\frac{\delta{\cal W}_k}{\delta\Sigma_i} -
L^i_k\Big(\frac{\delta{\cal W}_k}{\delta J}\Big) \,,\qquad i=1,2.
\eeq
From (\ref{Gcf1}) and (\ref{Gcf2}) follows that
\beq
\label{ConGkcf} \frac{\delta\Gamma_k}{\delta\Phi^A}=-J_A-
\Sigma_i\frac{\delta L^i_k(\Phi)}{\delta\Phi^A} \,,\qquad\quad
\frac{\delta\Gamma_k}{\delta F^i}=-\hbar\Sigma_i\,.
\eeq
Let us introduce
the full sets of fields ${\cal F}^{\cal A}$ and sources ${\cal
J}_{\cal A}$ according to
\beq {\cal F}^{\cal A}=(\Phi^A,\hbar F^i)
\,,\qquad {\cal J}_{\cal A}=(J_A,\hbar\Sigma_i).
\eeq

From the condition of solvability of equations (\ref{ConGkcf}) with
respect to the sources \ $J$ \ and \ $\Sigma$, it follows that
\beq
\frac{\de {\cal F}^{\cal
C}({\cal J})}{\de {\cal J}_{\cal B}}\,\,
\frac{\de_{l}{\cal J}_{\cal
A}({\cal F})}{\de{\cal F}^{\cal C}} \,=\,\de^{\cal B}_{\cal A}\,.
\eeq
One can express \ ${\cal J}_{\cal A}$ \ as a function of
the fields in the form
\beq
{\cal J}_{\cal A}
\,=\,\Big(-\frac{\de\Ga_k}{\de\Phi^A}\,-\,
\frac{\de\Ga_k}{\de F^i}\,\frac{\de L^i_k(\Phi)}{\de\Phi^A}
\,,\,\,-\,
\frac{\de\Ga_k}{\de F^i}\Big)
\eeq
and, therefore,
\beq
\frac{\delta_{l}{\cal J}_{\cal B}({\cal F})}{\delta{\cal F}^{\cal A}}
= -(G^{''}_k)_{{\cal A}{\cal B}}\,,\qquad
\frac{\delta {\cal F}^{\cal B}({\cal J})}{\delta {\cal J}_{\cal A}}=
-(G^{''-1}_k)^{{\cal A}{\cal B}}\,.
\eeq

Now we are in a position to derive the FRW equation for
$\Gamma_k=\Gamma_k(\Phi,K;F)$ and see that it gains more
complicated form due to the presence of composite fields.
The expression for the generating functional of connected Green's
functions can be obtained from (\ref{Zcf}),
\beq
\label{Wcf}
\exp \Big\{\frac{i}{\hbar}\,{\cal W}_k(J,K;\Sigma)\Big\}
\,=\,
\int {\cal D} \Phi^{'}\,
\exp \Big\{\frac{i}{\hbar}\,\big[S_{FP}(\Phi^{'})+J\Phi^{'}
+K{\hat s}\Phi^{'} + \Sigma L_k(\Phi^{'})\big]\Big\}\,.
\eeq
Taking into account Eq. (\ref{Gcf1}), the equation (\ref{Wcf})
in terms of \ $\Gamma_k$ \ reads
\beq
\nonumber
\label{Gcf}
&&\exp\Big\{\frac{i}{\hbar}\Big[\Gamma_k(\Phi,K;F)+J\Phi+
\Sigma\big(L_k(\Phi)+\hbar F\big)\Big]\Big\}
\\
\nonumber
&=& \int {\cal D} \Phi^{'}\,
\exp
\Big\{\frac{i}{\hbar}\,
\big[S_{FP}(\Phi^{'})+J\Phi^{'}+K{\hat s}\Phi^{'}
+ \Sigma L_k(\Phi^{'})\big]\Big\}\,.
\eeq
An equivalent form of this equation is
\beq
\nonumber
\label{Gcf-V2}
&&\exp\Big\{\frac{i}{\hbar}\Big[\Gamma_k(\Phi,K;F)+
\hbar\Sigma F\Big]\Big\}
\\
\nonumber
&=&\int {\cal D} \Phi^{'}\,
\exp \Big\{\frac{i}{\hbar}\,\big[S_{FP}(\Phi^{'})+J(\Phi^{'}-\Phi)+K{\hat s}\Phi^{'}
+ \Sigma \big(L_k(\Phi^{'})-L_k(\Phi)\big)\big]\Big\}\,.
\eeq
Making shift of the variables of integration in the functional
integral, \ $\Phi^{'}-\Phi=\varphi$, \ and using (\ref{ConGkcf}),
we obtain the equation for the effective action \
$\Gamma_k=\Ga_k(\Phi,K;F)$,
\beq
\nonumber
\exp \Big\{\frac{i}{\hbar}\Big[\Ga_k
\,-\, \frac{\de\Ga_k}{\de F}\,F
\Big]\Big\}
\,=\,
\int {\cal D}\varphi\;
\exp \bigg\{\frac{i}{\hbar}\,\Big[
S_{FP}(\Phi+\varphi)-
\frac{\de\Ga_k}{\delta\Phi}\,\varphi
\\
+\,K{\hat s}\,(\Phi+\varphi)
\,-\,\frac{1}{\hbar} \frac{\de\Ga_k}{\de F}\,
\Big(L_k(\Phi+\varphi)- L_k(\Phi)
\,-\,\frac{\de L_k(\Phi)}{\de\Phi}\,\varphi
\Big)\Big]\bigg\}\,.
\eeq
The solution of this equation in the tree-level approximation
has the form
\beq
\Gamma^{(0)}_k(\Phi,K;F) &=& S_{FP}(\Phi)+K{\hat s}\Phi\,,
\label{Gamma-0}
\eeq
which does not depend on the fields $F^i$ and parameter $k$.
The next step is to define loop corrections to (\ref{Gamma-0}),
so we assume that
\beq
\Gamma_k(\Phi,K;F)
&=& S_{FP}(\Phi)+K{\hat s}\Phi+\hbar{\bar \Gamma}_k(\Phi;F)\,.
\eeq
By taking into account the explicit structure of regulator functions
we obtain the equation which can be, for example, a basis for deriving
the loop expansion of \ ${\bar\Gamma}_k={\bar\Gamma}_k(\Phi;F)$,
\beq
\nonumber
\exp \bigg\{ i\Big[ {\bar\Ga}_k
\,-\, \frac{\de{\bar\Ga}_k}{\de F}\,F\Big]\bigg\}
&=&
\int {\cal D}\varphi\;
\exp \bigg\{\frac{i}{\hbar}\,\Big[
S_{FP}(\Phi+\varphi)
- S_{FP}(\Phi)-\frac{\delta S_{FP}(\Phi)}{\delta\Phi}\varphi
\\
&-&
\hbar\,\frac{\de{\bar\Ga}_k}{\delta\Phi}\,\varphi
-\frac{\de{\bar\Ga}_k}{\de F}\,
L_k(\varphi)\Big]\bigg\}\,.
\eeq

Using the properties of the Legendre transformation, one can
arrive at the relation
\beq
\pa_t{\cal W}_k\,=\,\pa_t \Ga_k
\,-\,\frac{1}{\hbar}\,\frac{\de\Ga_k}{\de F^i}\,\,
\pa_tL^i_k(\Phi)\,,
\eeq
where the derivatives $\pa_t$ are  calculated with respect
to the explicit dependence of \ $\Ga_k$, \ ${\cal W}_k$ and
$L^i_k(\Phi)$ on the regulator parameter $k$.

Finally, the FRG flow equation, in terms of the functional
$\Gamma_k$, is cast into the form
\beq
\pa_t \Ga_k
&=&
-\,i\,\Big\{\frac{1}{2}\,\frac{\de\Ga_k}{\de F^1}
\,\,\pa_t(R_{k,A})^{ab}_{\mu\nu} \,\,
\big(G^{''-1}_k\big)^{(a\mu)(b\nu)}
\,+\,
\frac{\de\Ga_k}{\de F^2}\,\pa_t(R_{k,gh})\,
\big(G^{''-1}_k\big)^{ab}\Big\}
\nonumber
\\
\label{RGfcf}
&=&
- \,\frac{i}{2}\,
{\rm Tr}\Big\{\frac{\delta\Gamma_k}{\delta F^1}\,
\pa_t(R_{k,A}) (G^{''-1}_k)\Big\}_A
\,+\, i\,{\rm Tr} \Big\{\frac{\de\Ga_k}{\de F^2}\,\,
\pa_t(R_{k,gh})\,\,(G^{''-1}_k)\Big\}_C\,,
\eeq
when we used usual traces in the sectors of vector $A^{a\mu}$
and ghost $C^a$ fields while the Grassmann parity of quantum
fields is taken into account explicitly.

Consider now a consequence of the BRST invariance of
\ $S_{FP}$ \ for the generating functional (\ref{Zcf}).
Making use of the change of variables (\ref{BRSTtr}) in
(\ref{Zcf}) and taking into account the nilpotency of
BRST transformation (\ref{BRSTnil}), we arrive at the
identity
\beq
\nonumber
&&
\int {\cal D}\Phi\;\Big\{
J_A{\hat s}\,\Phi^A
\,+\,
\Si_1(R_{k,A})^{ab}_{\mu\nu}\,A^{b\nu}\,{\hat s}A^{a\mu}
\,+\,
\Si_2 {\bar C}^a(R_{k,gh})^{ab}\,{\hat s}C^a
\\
\label{WIZcf}
&&
\qquad
-\,\Si_2(R_{k,gh})^{ab}\,C^b\,{\hat s}{\bar C}^a\Big\}
\,\exp \Big\{
\frac{i}{\hbar}\big[S_{FP}({\Phi})+J\Phi
+K{\hat s}\Phi+ \Sigma L_k\Big] \Big\}
\,\equiv\, 0\,.
\eeq
The same equation can be written in terms of the
generating functional (\ref{Zcf}),
\beq
\label{WIZcf1}
J_A\frac{\de {\cal Z}_k}{\de K_A}
+ \frac{\hbar}{i}\Big\{
\Sigma_1(R_{k,A})^{ab}_{\mu\nu}
\frac{\delta^2 {\cal Z}_k}{\delta j^b_{\nu}\delta K^a_{\mu}}
+ \Sigma_2(R_{k,gh})^{ab}
\frac{\delta^2 {\cal Z}_k}{\delta \eta^a\delta {\bar L}^a}
- \Sigma_2(R_{k,gh})^{ab}
\frac{\delta^2 {\cal Z}_k}{\delta {\bar\eta}^b\delta L^a}\Big\}
\,\equiv\, 0\,.
\eeq
The identity (\ref{WIZcf1}) can be considered as the generalized
 Slavnov-Taylor identity in the presence of composite fields.
In the limits of \ $\Sigma_1(x)=\Sigma_2(x)=0$ \ and
\ $\Sigma_1(x)=\Sigma_2(x)=1$ this identity coincides
with the previously considered ideintities (\ref{WIZJKI}) and
(\ref{STSk}), respectively.

Let us explore the gauge dependence of generating functional
of Green's functions in the presence of composite fields,
(\ref{Zcf}).  As before, one can consider an infinitesimal
variation of gauge function, \
$\chi^a\rightarrow \chi^a+\delta\chi^a$. \
Taking into account the results obtained in Sect. 2,
we arrive at the following variation of
${\cal Z}_k={\cal Z}_k(J,K;\Sigma)$:
\beq
\de{\cal Z}_k
&=&
\frac{i}{\hbar} \int {\cal D}\Phi\,\,
\frac{\de\,\de\psi}{\de\Phi^A}\,\,
\frac{\delta (K{\hat s}\Phi)}{\delta K_A}
\exp \Big\{
\frac{i}{\hbar}\big[
S_{FP}({\Phi})+J\Phi +K{\hat s}\Phi+\Sigma L_k\big]\Big\}\,,
\label{6.22}
\eeq
where the Grassmann-odd functional
\ $\de\psi={\bar C}^a\delta\chi^a$ \ has been introduced.
Starting from the identity
\beq
\label{EQU3-bis}
\int {\cal D}\Phi\,\,
\frac{\delta}{\delta\Phi^A}\,
\bigg[
\de\psi\,
\frac{\delta (K{\hat s}\Phi)}{\delta K_A}
\,\exp \Big\{\frac{i}{\hbar}\Big[
S_{FP}({\Phi})+J\Phi +K{\hat
s}\Phi+\Sigma L_k\Big]\Big\}\bigg]\,=\,0\,,
\eeq
one can derive the relation
\beq
\label{EQU4-bis}
&&
\int {\cal D}\Phi\,\,
\,\frac{\de\,\de\psi}{\de\Phi^A}\,
\frac{\de (K{\hat s}\Phi)}{\de K_A}\,\,
\exp \Big\{ \frac{i}{\hbar}\big[
S_{FP}({\Phi})+J\Phi +K{\hat s}\Phi+\Si L_k\Big]\Big\}
\\
&&
\nonumber
\;
=\,-\,\frac{i}{\hbar}\int {\cal D}\Phi
\; \de\psi\Big(J_A + \Si\,\frac{\de L_k}{\de \Phi^A}\Big)
\,\frac{\de (K{\hat s}\Phi)}{\de K_A}\,
\exp \Big\{  \frac{i}{\hbar}
\big[S_{FP}(\Phi) + J\Phi + K{\hat s}\Phi + \Si L_k\big]
\Big\}\,.
\eeq
Using the last formula together with Eq. (\ref{6.22}), one can
show that the gauge dependence of ${\cal Z}_k$ is described by
the equation
\beq
\nonumber
\de{\cal Z}_k
 &=& -\,\Big(\frac{i}{\hbar}\Big)^2
 \int {\cal D} \Phi\,\de\psi\,
 \Big(J_A + \Si\,\frac{\de L_k}{\de\Phi^A}\Big)
 \,\frac{\de (K{\hat s}\Phi)}{\de K_A}\,\times
\\
\label{dZk}
&\times&
\exp \Big\{ \frac{i}{\hbar}\big[
S_{FP}({\Phi}) + J\Phi + K{\hat s}\Phi + \Si L_k\big]
\Big\}\,.
\eeq
Taking into account the explicit structure of regulator functions
$L^i_k(\Phi)$ in Eqs. (\ref{L1}) and (\ref{L2}), one can rewrite
the second term in the integrand of (\ref{dZk}) as
\beq
\nonumber
&& \Si\,\frac{\delta L_k}{\delta\Phi^A}\,
\frac{\delta (K{\hat s}\Phi)}{\delta K_A}
\,=\,
\Sigma_1 A^{a\mu}(R_{k,A})^{ab}_{\mu\nu}\,
\frac{\delta (K{\hat s}\Phi)}{\delta j^b_{\nu}}
\\
&& +\,\Sigma_2\,{\bar C}^a(R_{k,gh})^{ab}\,
\frac{\delta (K{\hat s}\Phi)}{\delta {\bar L}^b}
\,-\,\Sigma_2\,C^a(R_{k,gh})^{ba}
\,\frac{\delta (K{\hat s}\Phi)}{\delta L^b}\,.
\eeq
Using this relation, we obtain the final equation, describing
the gauge dependence of the generating functional
\ ${\cal Z}_k={\cal Z}_k(J,K;\Sigma)$,
\beq
\nonumber
\delta{\cal Z}_k
&=&
\frac{i}{\hbar}\,J_A\,\frac{\delta}{\delta K_A}\,
\delta\psi\,\Big(\frac{\hbar}{i}\frac{\delta}{\delta J}\Big)
{\cal Z}_k
\,+\, \Big\{
\Si_1(R_{k,A})^{ab}_{\mu\nu}\,
\frac{\de^2{\cal Z}_k}{\de j^a_{\mu}\delta j^b_{\nu}}
\\
\label{varZcf}
&+&
\Si_2(R_{k,gh})^{ab}\,
\frac{\delta^2{\cal Z}_k}{\delta \eta^a\delta {\bar L}^b}
\,-\,\Sigma_2(R_{k,gh})^{ba}\,
\frac{\de^2{\cal Z}_k}{\de {\bar\eta}^a\de L^b}\Big\}
\,\de\psi\Big(\frac{\hbar}{i}\frac{\de}{\de J}\Big)\,
{\cal Z}_k\,.
\eeq
In the limits \ $\Sigma_1(x)=\Sigma_2(x)=0$ \
and
\ $\Sigma_1(x)=\Sigma_2(x)=1$, \ this equation coincides
with Eqs. (\ref{varZ16}) and (\ref{varZkf}), respectively.

In terms of generating functional of connected Green's
functions,
\ ${\cal W}_k={\cal W}_k(J,K)$, the equation (\ref{varZcf})
can be written as
\beq
\nonumber
\delta{\cal W}_k
&=&
{\cal J}_A\,\frac{\de}{\de K_A}\,
\de\psi\Big(\frac{\de{\cal W}_k}{\de J}
+ \frac{\hbar}{i}\,\frac{\de}{\de J}\Big)
\,-\,i\hbar
\Big\{
\Sigma_1(R_{k,A})^{ab}_{\mu\nu}\,
\frac{\delta^2}{\delta j^a_{\mu}\delta j^b_{\nu}}
\\
&+&
\Sigma_2(R_{k,gh})^{ab}
\frac{\delta^2}{\delta \eta^a\delta {\bar L}^b}
\,-\,\Si_2(R_{k,gh})^{ba}\,
\frac{\de^2}{\de {\bar\eta}^a\,\de L^b}\Big\}
\,\de\psi\Big(\frac{\delta{\cal W}_k}{\delta J}+
\frac{\hbar}{i}\frac{\delta}{\delta J}\Big)\,,
\label{varWkcf}
\eeq
where the identity (\ref{WIWkJ}) was used. If we
introduce the notations
\beq
L^i_{k,A}(\Phi)=\frac{\pa L^i_k(\Phi)}{\pa \Phi^A}\,,
\qquad i=1,2\,,
\eeq
the equation (\ref{varWkcf}) can be cast in the
compact form,
\beq
\delta{\cal W}_k
&=&
\Big\{J_A\,-\,i\hbar\, \Si_i\,L^i_{k,A}
\Big(\frac{\hbar}{i}\frac{\delta}{\delta J}\Big)
\Big\}\,\frac{\de}{\de K_A}\,
\de\psi\Big(\frac{\delta{\cal W}_k}{\delta J}+
\frac{\hbar}{i}\frac{\delta}{\delta J}\Big)\,.
\label{6.30}
\eeq
Finally, the gauge dependence of the effective average
action follows from (\ref{6.30}) after one performs the
Legendre transform. It is described by the equation
\beq
\label{varGkcf}
\delta \Gamma_k
&=& -
\,\frac{\de\Ga_k}{\de\Phi^A}\,
\frac{\de}{\de K_A} \de\psi\big({\hat \Phi}\big)
\,-\,\frac{1}{\hbar}\,\frac{\de\Ga_k}{\de F^i}\,
L^i_{k,A}\,(\Phi - {\hat\Phi})\,\frac{\de}{\de K_A}\,
\delta\psi\big({\hat \Phi}\big)\,.
\eeq
Let us define the mass-shell of the quantum theory
by the equations
\beq
\label{newosh}
\frac{\de\Ga_k}{\de\Phi^A}=0
\,,\qquad
\frac{\de\Ga_k}{\de F^i}\,=\,0\,.
\eeq
Then from (\ref{varGkcf}) immediately follows the gauge
independence of the effective action
\ $\Ga_k=\Ga_k(\Phi,K;F)$ \ on-shell. Namely, when the
relations (\ref{newosh}) are satisfied, we have
$\de \Ga_k=0$. Moreover, all physical quantities
calculated on the basis of the modified version of the
effective average action do not depend on gauge and on
the parameter $k$. The last feature is common for the
renormalization group based on the abstract scale parameters
(such as cut-off, $k$ or $\mu$), which require an additional
identification of scale to be applied to one or another physical
problem. For example, we know that the $S$-matrix elements in
usual Quantum Field Theory do not depend on $\mu$ in the Minimal
Subtraction scheme of renormalization. This does not mean that
there is no running, of course (see \cite{DCCR} for detailed
discussion of this issue). In our case, after the evaluation
of $\Gamma_k$ is completed in a given approximation (including,
perhaps, the truncation scheme) one has to identify $k$ with
some physical quantity and only after that go on-shell and
calculate physical quantities, such as $S$-matrix
elements\footnote{In most complicated cases, such
as Quantum (or semiclassical) Gravity, when direct comparison
with the physical renormalization scheme (like momentum
subtraction) is not possible, the scale-setting is tricky,
but still admits a regular procedure which works well in
different physical situations \cite{babic,DomStef}.}.
\vskip 2mm

\section{Conclusions and final discussions}

\noindent
We investigated, in much more details than it was done before,
the problem of gauge dependence in the functional renormalization
group (FRG) approach. The consideration was performed for the
generating functional of the Green functions and effective action,
but the main target was the universality of the definition of
$S$-matrix and, more general, on-shell gauge dependence of
the effective action.

The regulator functions which are introduced in the FRG formalism
to model the behavior of exact Green's functions lead to the
breakdown of the BRST symmetry. As a result, the effective
average action depends on the choice of gauge fixing
even on-shell. The situation is qualitatively similar to the
Gribov-Zwanziger theory \cite{Gribov,Zwanziger1,Zwanziger2},
where the restrictions on the domain of integration in functional
integral, which is due to the Gribov horizon, violates the BRST
symmetry and consequently leads to on-shell gauge dependence of
the effective average action \cite{LLR,LRR}. As another example
of this sort we can mention the situation with the theory
possessing global supersymmetry (modified BRST symmetry)
\cite{Lav-MPLA-2012}. When the nilpotency of the global
supersymmetry is violated, one meets a problem of gauge
dependence for the relevant physical quantities. One can
suppose that the gauge fixing dependence will affect both
perturbative and non-perturbative results within the FRG
approach, because it concerns the effective average action
before the truncation scheme is chosen. Of course, one can
hope that the physical relevance of the results will be
restored in the non-perturbative regime, but there are no
explicit reasons to see why should this occur.

One can consider the situation with gauge dependence of $S$-matrix
and on-shell effective action in two different ways. The first one
implies that we consider the effective average action as an
approximation to the real effective action in a given theory.
As far as the last is gauge independent on-shell and leads to the
well-defined physical predictions, one may think that the effective
average action produces an approximation to the invariant quantities
in a given gauge, which should be taken as granted. This kind of
consideration is, in some sense, the unique option if we do not
invent an alternative gauge-invariant formulation of the FRG
approach. However, it is obvious that the second way, that
is constructing a gauge-invariant formulation, would mean
a much better approximation.

As a first step forward in formulating the gauge-invariant version
of the FRG, we propose the new formulation of the theory with
cut-off dependent regulator functions. It was shown that if these
functions are introduced by means of the special composite fields,
the BRST symmetry is preserved at quantum level, vacuum functional
does not depend on the choice of gauge and, finally, the new theory
is free from the on-shell gauge dependence. It would be very
interesting to compare this new approach to the standard one.
For example, it looks interesting to verify the gauge independence
for the approximate higher-loop $\be$-function of the theory
in the new and traditional approaches. The derivation of such
a $\be$-functions and, in general, the loop expansion within the
FRG theory with composite fields, lies beyond the scope of the
present work. Indeed, we expect to deal with these and other
related subjects in the next publications.

\section*{Acknowledgments}

\noindent I.Sh. is grateful to Roberto Percacci and Andreas Wipf for
stimulating discussions on the functional renormalization group. The
present work was completed during the visit of P.M.L. to Juiz de
Fora in the framework of the program PVE of CAPES and we wish to
acknowledge CAPES for supporting this visit. The work of P.M.L. is
also partially supported by the LRSS grant 224.2012.2, as well as by
the RFBR grant 12-02-00121, by the Ministry of Education and Science
of Russian Federation, project 14.B37.21.0774 and by the
RFBR-Ukraine grant 13-02-90430. I.Sh. is thankful to CNPq, CAPES,
FAPEMIG and ICTP for partial support of his work.
\vskip 6mm


\end{document}